\RequirePackage{amsmath}
\documentclass[nolinenum]{jfp} 
\usepackage{tikz-cd}
\pgfrealjobname{main}
\usepackage{quoting}
\usepackage{bigfoot} 
\usepackage{hyperref}
\usepackage[utf8]{inputenc}
\usepackage{epstopdf}

\makeatletter
\@ifundefined{lhs2tex.lhs2tex.sty.read}%
  {\@namedef{lhs2tex.lhs2tex.sty.read}{}%
   \newcommand\SkipToFmtEnd{}%
   \newcommand\EndFmtInput{}%
   \long\def\SkipToFmtEnd#1\EndFmtInput{}%
  }\SkipToFmtEnd

\newcommand\ReadOnlyOnce[1]{\@ifundefined{#1}{\@namedef{#1}{}}\SkipToFmtEnd}
\usepackage{amstext}
\usepackage{amssymb}
\usepackage{stmaryrd}
\DeclareFontFamily{OT1}{cmtex}{}
\DeclareFontShape{OT1}{cmtex}{m}{n}
  {<5><6><7><8>cmtex8
   <9>cmtex9
   <10><10.95><12><14.4><17.28><20.74><24.88>cmtex10}{}
\DeclareFontShape{OT1}{cmtex}{m}{it}
  {<-> ssub * cmtt/m/it}{}

\DeclareFontShape{OT1}{cmtt}{bx}{n}
  {<5><6><7><8>cmtt8
   <9>cmbtt9
   <10><10.95><12><14.4><17.28><20.74><24.88>cmbtt10}{}
\DeclareFontShape{OT1}{cmtex}{bx}{n}
  {<-> ssub * cmtt/bx/n}{}

\newcommand{\Conid}[1]{\mathit{#1}}
\newcommand{\Varid}[1]{\mathit{#1}}
\newcommand{\anonymous}{\kern0.06em \vbox{\hrule\@width.5em}}

\renewcommand{\leq}{\leqslant}
\renewcommand{\geq}{\geqslant}
\usepackage{polytable}

\@ifundefined{mathindent}%
  {\newdimen\mathindent\mathindent\leftmargini}%
  {}%

\def\resethooks{%
  \global\let\SaveRestoreHook\empty
  \global\let\ColumnHook\empty}
\newcommand*{\savecolumns}[1][default]%
  {\g@addto@macro\SaveRestoreHook{\savecolumns[#1]}}
\newcommand*{\restorecolumns}[1][default]%
  {\g@addto@macro\SaveRestoreHook{\restorecolumns[#1]}}
\newcommand*{\aligncolumn}[2]%
  {\g@addto@macro\ColumnHook{\column{#1}{#2}}}

\resethooks

\newcommand{\onelinecommentchars}{\quad-{}- }
\newcommand{\commentbeginchars}{\enskip\{-}
\newcommand{\commentendchars}{-\}\enskip}

\newcommand{\visiblecomments}{%
  \let\onelinecomment=\onelinecommentchars
  \let\commentbegin=\commentbeginchars
  \let\commentend=\commentendchars}

\newcommand{\invisiblecomments}{%
  \let\onelinecomment=\empty
  \let\commentbegin=\empty
  \let\commentend=\empty}

\visiblecomments

\newlength{\blanklineskip}
\newcommand{\hsindent}[1]{\quad}
\let\hspre\empty
\let\hspost\empty

\EndFmtInput
\makeatother
\ReadOnlyOnce{polycode.fmt}%
\makeatletter

\newcommand{\hsnewpar}[1]%
  {{\parskip=0pt\parindent=0pt\par\vskip #1\noindent}}

\newcommand{\hscodestyle}{}

\newcommand{\sethscode}[1]%
  {\expandafter\let\expandafter\hscode\csname #1\endcsname
   \expandafter\let\expandafter\endhscode\csname end#1\endcsname}

  {\par\noindent
   \advance\leftskip\mathindent
   \hscodestyle
   \let\\=\@normalcr
   \let\hspre\(\let\hspost\)%
   \pboxed}%
  {\endpboxed\)%
   \par\noindent
   \ignorespacesafterend}

  {\hsnewpar\abovedisplayskip
   \advance\leftskip\mathindent
   \hscodestyle
   \let\hspre\(\let\hspost\)%
   \pboxed}%
  {\endpboxed%
   \hsnewpar\belowdisplayskip
   \ignorespacesafterend}

  {\hsnewpar\abovedisplayskip
   \advance\leftskip\mathindent
   \hscodestyle
   \let\\=\@normalcr
   \(\pboxed}%
  {\endpboxed\)%
   \hsnewpar\belowdisplayskip
   \ignorespacesafterend}

\newcommand{\plainhs}{\sethscode{plainhscode}}

\plainhs

  {\hsnewpar\abovedisplayskip
   \advance\leftskip\mathindent
   \hscodestyle
   \let\\=\@normalcr
   \(\parray}%
  {\endparray\)%
   \hsnewpar\belowdisplayskip
   \ignorespacesafterend}

  {\parray}{\endparray}

  {\(\parray}{\endparray\)}

\def\codeframewidth{\arrayrulewidth}
\RequirePackage{calc}

  {\parskip=\abovedisplayskip\par\noindent
   \hscodestyle
   \arrayrulewidth=\codeframewidth
   \tabular{@{}|p{\linewidth-2\arraycolsep-2\arrayrulewidth-2pt}|@{}}%
   \hline\framedhslinecorrect\\{-1.5ex}%
   \let\endoflinesave=\\
   \let\\=\@normalcr
   \(\pboxed}%
  {\endpboxed\)%
   \framedhslinecorrect\endoflinesave{.5ex}\hline
   \endtabular
   \parskip=\belowdisplayskip\par\noindent
   \ignorespacesafterend}

\newcommand{\framedhslinecorrect}[2]%
  {#1[#2]}

  {\(\def\column##1##2{}%
   \let\>\undefined\let\<\undefined\let\\\undefined
   \newcommand\>[1][]{}\newcommand\<[1][]{}\newcommand\\[1][]{}%
   \def\fromto##1##2##3{##3}%
   }{\) }%

\newenvironment{joincode}%
  {\let\orighscode=\hscode
   \let\origendhscode=\endhscode
   \def\endhscode{\def\hscode{\endgroup\def\@currenvir{hscode}\\}\begingroup}
   \orighscode\def\hscode{\endgroup\def\@currenvir{hscode}}}%
  {\origendhscode
   \global\let\hscode=\orighscode
   \global\let\endhscode=\origendhscode}%

\makeatother
\EndFmtInput
\def\doubleequals{\mathrel{\unitlength 0.01em
  \begin{picture}(78,40)
    \put(7,34){\line(1,0){25}} \put(45,34){\line(1,0){25}}
    \put(7,14){\line(1,0){25}} \put(45,14){\line(1,0){25}}
  \end{picture}}}

\newcommand{\fixatill}[3]{#3^{#1}_{#2}}

\newcommand{\majbrace}[4]{\underbrace{#1,#2,#3}_{#4}}

\def\commentbegin{\quad\{\ }
\def\commentend{\}}

\newcommand{\T}{\ensuremath{\mathbf{1}}}
\newcommand{\F}{\ensuremath{\mathbf{0}}}

\newcommand{\val}[1]{x_{#1}}

\providecommand{\TODO}[2][?]{\marginpar{\raggedright \tiny TODO: #2}}

\newcommand{\genAlgNode}[3]{\ensuremath{#2 \mapsto #3}}

\usepackage{nicefrac}

\usepackage{comment}                
\usepackage{moreverb}								
\usepackage{textcomp}								
\usepackage{lmodern}								
\usepackage{helvet}									
\usepackage[T1]{fontenc}						
\usepackage[english]{babel}					
\usepackage[utf8]{inputenc}					
\usepackage{graphicx}								
\counterwithin{figure}{section}
\counterwithin{table}{section}
 
\usepackage{amsmath, amssymb, amsfonts,amsthm,thmtools}
\usepackage{csquotes}
\usepackage{subcaption} 
\usepackage{siunitx}
\usepackage{hyperref}
\hypersetup{
    colorlinks=true,
    linkcolor=blue,
    filecolor=magenta,      
    urlcolor=blue,
}
\usepackage[shortlabels]{enumitem}

\usepackage{tikz,forest}
\usepackage[justification=centering]{caption}
\usetikzlibrary{trees}
\usetikzlibrary{arrows.meta,tikzmark}
\usetikzlibrary{graphs, graphs.standard, quotes}
\tikzset{
  treenode/.style = {shape=rectangle, rounded corners,
                     draw, align=center, 
                     minimum height=2ex, text depth=0.25ex,
                     top color=white, bottom color=blue!20},
  root/.style     = {treenode, font=\normalsize\rmfamily},
  env/.style      = {treenode, font=\ttfamily\normalsize},
  every label/.append style = {label distance=15pt, inner sep=1pt, 
                             align=left, font=\tiny}, 
}

\begin{document}

\journaltitle{JFP}
\cpr{The Author(s),}
\doival{10.1017/xxxxx}

\lefttitle{Jansson and Jansson}
\righttitle{Level-\ensuremath{\Varid{p}}-complexity of Boolean Functions}
\totalpg{\pageref{lastpage01}}
\jnlDoiYr{2023}

\title{Level-\ensuremath{\Varid{p}}-complexity of Boolean Functions using Thinning, Memoization, and Polynomials}
\begin{authgrp}
\author{JULIA JANSSON}
\affiliation{Chalmers University of Technology and University of Gothenburg, Göteborg, Sweden.
  (\email{juljans@chalmers.se})}
\author{PATRIK JANSSON}
\affiliation{Chalmers University of Technology and University of Gothenburg, Göteborg, Sweden.
  (\email{patrikj@chalmers.se})}
\end{authgrp}

\begin{abstract}
  This paper describes a purely functional library for computing level-$p$-complexity of Boolean functions, and applies it to two-level iterated majority.
  Boolean functions are simply functions from $n$ bits to one bit, and they can describe digital circuits, voting systems, etc.
  An example of a Boolean function is majority, which returns the value that has majority among the $n$ input bits for odd $n$.
  The complexity of a Boolean function $f$ measures the \emph{cost} of evaluating it: how many bits of the input are needed to be certain about the result of $f$.
  There are many competing complexity measures but we focus on level-$p$-complexity --- a function of the probability $p$ that a bit is 1.
  The level-$p$-complexity \(D_p(f)\) is the minimum expected cost when the input bits are independent and identically distributed with Bernoulli($p$) distribution.
  We specify the problem as choosing the minimum expected cost of all possible decision trees --- which directly translates to a clearly correct, but very inefficient implementation.
  The library uses thinning and memoization for efficiency and type classes for separation of concerns.
  The complexity is represented using (sets of) polynomials, and the order relation used for thinning is implemented using polynomial factorisation and root-counting.
  Finally we compute the complexity for two-level iterated majority and improve on an earlier result by J.~Jansson.
\end{abstract}

\maketitle


\setlength{\mathindent}{1em}
\newcommand{\fixlengths}{\setlength{\abovedisplayskip}{6pt plus 1pt minus 1pt}\setlength{\belowdisplayskip}{6pt plus 1pt minus 1pt}}
\renewcommand{\hscodestyle}{\small\fixlengths}
\fixlengths

\section{Introduction}
Imagine a voting system with yes/no options, for example direct democracy, indirect democracy, or dictatorship.
How much information of the votes do we need until we can conclude the outcome of the election?
For dictatorship, we only need the information of the dictator as he or she has all the power, but for a democratic majority we need at least half the votes.
Depending on the order in which we find out what the votes are we might need all of them before we can conclude the result.
More generally, this question is about complexity of Boolean functions which is application area of this paper.

Boolean functions are wide-spread in mathematics and computer science and can describe yes-no voter systems, hardware circuits, and predicates \citep{o2014analysis, knuth2011art}.
A Boolean function is a function from \ensuremath{\Varid{n}} bits to one bit, for example majority (\ensuremath{\Varid{maj}_{\Varid{n}}}), which returns the value that has majority among the \ensuremath{\Varid{n}} inputs.
In the context of voting systems, the next subsection gives an example of a Boolean function called iterated majority.

\subsection{Vote counting example: iterated majority}
In US elections a presidential candidate can lose even if they win the popular vote.
One reason for this is that the outcome is not directly determined by the majority, but rather majority iterated two times.%
\footnote{The actual presidential election is a direct majority vote among the electors who are not formally bound by their states' outcome.}
Our running example is a very much simplified case: consider 3 states with 3 voters in each.
\[\majbrace{\majbrace{\val{(0,0)}}{\val{(0,1)}}{\val{(0,2)}}{m_0 = \ensuremath{\Varid{maj}_{3}\;(\mathbin{...})}}}
    {\majbrace{\val{(1,0)}}{\val{(1,1)}}{\val{(1,2)}}{m_1 = \ensuremath{\Varid{maj}_{3}\;(\mathbin{...})}}}
    {\majbrace{\val{(2,0)}}{\val{(2,1)}}{\val{(2,2)}}{m_2 = \ensuremath{\Varid{maj}_{3}\;(\mathbin{...})}}}{\ensuremath{\Varid{maj}_{3}}(m_0,m_1,m_2)}
\]
We first compute the majority $m_i$ in each ``state'', and then the majority of $m_0$, $m_1$, and $m_2$.
For example we see below $\F,\T,\F$ which gives $m_0 =  \F$, then $\T,\F,\T$ which gives $m_1 = \T$, and $\F,\T,\F$ again which gives $m_2 = \F$.
The final majority is \F:
\[\majbrace{\majbrace{\F}{\T}{\F}{m_0 = \F}}
    {\majbrace{\T}{\F}{\T}{m_1 = \T}}
    {\majbrace{\F}{\T}{\F}{m_2 = \F}}{\ensuremath{\Varid{maj}_{3}} = \F}
\]
\noindent
But if we switch the first and 8th bit (perhaps through
gerrymandering) we get another result (with the changed bits
underlined and marked in red):
\[\majbrace{\majbrace{\underline{\textcolor{red}{\T}}}{\T}{\F}{m_0 = \underline{\textcolor{red}{\T}}}}
    {\majbrace{\T}{\F}{\T}{m_1 = \T}}
    {\majbrace{\F}{\underline{\textcolor{red}{\F}}}{\F}{m_2 = \F}}{\ensuremath{\Varid{maj}_{3}} = \underline{\textcolor{red}{\T}}}
\]
This changes $m_0$ from \F{} to \T{} without affecting $m_1$, or $m_2$.
But now the two-level majority is changed to \T, just from the switch of two bits.
Both examples have four \T{}'s  and five \F{}'s  but the result is different based on the positioning of the bits.
In our case the two-level majority is \T{} even though there are fewer \T{}'s  than \F{}'s.
This means that the \F{}'s ``lose'' even though they won the ``popular vote''.

\subsection{Cost and complexity}
The field of computational complexity is about “how much” computation is necessary and sufficient to perform certain computational tasks.
For example, given a computational problem it tries to establish tight upper and lower bounds on the length of the computation (or on other resources, like space).
Unfortunately, for many practically relevant computational problems no tight bounds are known.
In our case we study one of the simplest models of computation: the decision tree.
We are interested in the cost of evaluating Boolean functions and we use binary decision trees to describe the evaluation order of Boolean functions.
The depth of the decision tree corresponds to the number of votes needed to know the outcome for certain.
This is called deterministic complexity.
Another well-known notion is randomized complexity, and the randomized complexity bounds of iterated majority have been studied in \cite{landau2006lower}, \cite{leonardos2013improved} and \cite{magniez2016improved}.
Iterated majority on two levels corresponds to the Boolean function for US elections as described above.
We are particularly interested in this function due to its symmetry and simplicity, but still the complexity is non-trivial.

Diving into the literature for complexity of Boolean functions we find many different measures.
Relevant concepts are certificate complexity, degree of a Boolean function, and communication complexity \citep{buhrman2002complexity}.
Complexity measures related specifically to circuits are circuit complexity, additive, and multiplicative complexity \citep{wegener1987complexity}.
Considering Boolean computation in practice we have combinational complexity which is the length of the shortest Boolean chain computing it \citep{knuth2011art}.
Thus, there are many competing complexity measures but we focus on level-\ensuremath{\Varid{p}}-complexity --- a function of the probability $p$ that a bit is \T{} \citep{garban2014noise}.
We assume that the bits are independent and identically Bernoulli-distributed with parameter $p \in [0,1]$.
Then, for each Boolean function \ensuremath{\Varid{f}} and probability \ensuremath{\Varid{p}}, we get the level-\ensuremath{\Varid{p}}-complexity by minimizing the expected cost over all decision trees.
The level-\ensuremath{\Varid{p}}-complexity is a piecewise polynomial function of \ensuremath{\Varid{p}} and has many interesting properties \citep{jansson2022level}.

%

\subsection{Contributions}
\label{sec:aim}
This paper presents a purely functional library for computing level-\ensuremath{\Varid{p}}-complexity of Boolean functions in general, and for \ensuremath{\Varid{maj}_{3}^2} in particular.
The level-\ensuremath{\Varid{p}}-complexity of \ensuremath{\Varid{maj}_{3}^2} was conjectured in \citet{jansson2022level}, but could not be proven because it was hard to generate all possible decision trees.
This paper fills that gap, by showing that the conjecture is false and by computing the true level-\ensuremath{\Varid{p}}-complexity of \ensuremath{\Varid{maj}_{3}^2}.

The strength of our implementation is that it can calculate the level-$p$-complexity for Boolean functions quickly and correctly, compared to tedious calculations by hand.
Our specification uses exhaustive search and considers all possible candidates (decision trees).
Some partial candidates dominate (many) others, which may be discarded.
Thinning \citep{bird_gibbons_2020} is an algorithmic design technique which maintains a small set of partial candidates which provably dominate all other candidates.
We hope that one contribution of this paper is an interesting example of how a combination of algorithmic techniques can be used to make the intractable tractable.
The code in this paper is available on GitHub\footnote{The paper repository is at \url{https://github.com/juliajansson/BoFunComplexity}.} and uses packages from \citet{JanssonIonescuBernardyDSLsofMathBook2022}.
The implementation is in Haskell but should work also in other languages, and parts of it has been reproduced in Agda to check some of the stronger invariants.
The choice of Haskell for the implementation is due to its strong compiler and the availability of libraries for BDDs, memoization, and polynomials.
% ...
\subsection{Motivation}
\begin{figure}[tbp]
        \centering
\begin{tikzpicture}
\node at (0,0){\includegraphics[width=0.7\textwidth]{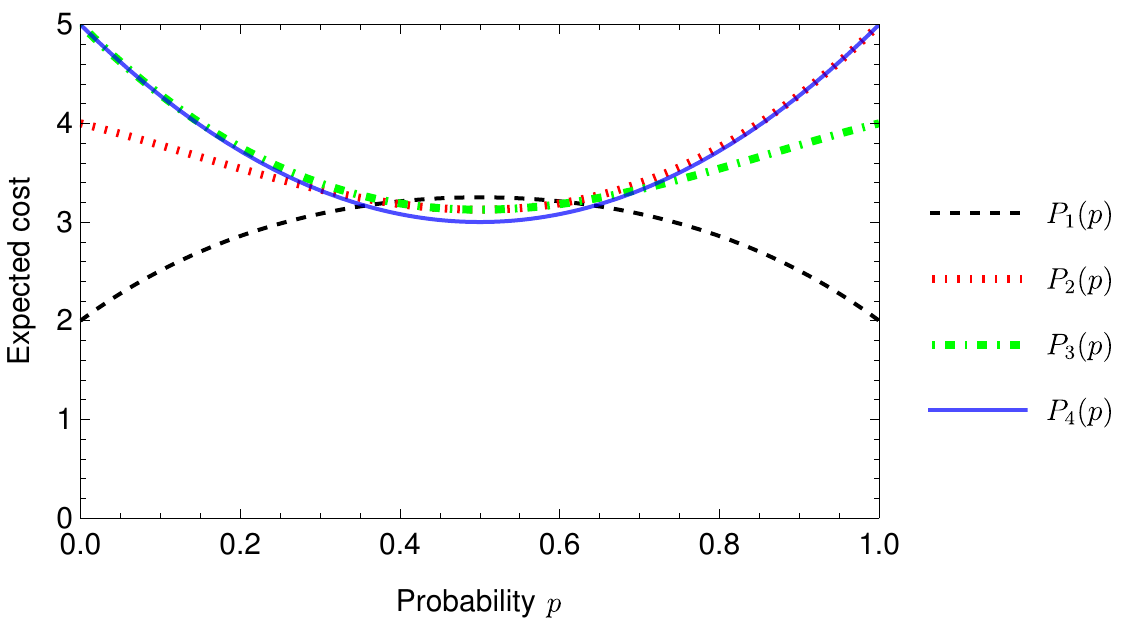}};
\node at (-0.5,-0.5){%
  \begin{minipage}{0.3\textwidth}%
    \tiny
    \begin{hscode}\SaveRestoreHook
\column{B}{@{}>{\hspre}l<{\hspost}@{}}%
\column{7}{@{}>{\hspre}l<{\hspost}@{}}%
\column{9}{@{}>{\hspre}l<{\hspost}@{}}%
\column{E}{@{}>{\hspre}l<{\hspost}@{}}%
\>[7]{}\Varid{sim}_{5}\;[\mskip1.5mu \Varid{x}_{0},\Varid{x}_{1},\Varid{x}_{2},\Varid{x}_{3},\Varid{x}_{4}\mskip1.5mu]\mathrel{=}{}\<[E]%
\\
\>[7]{}\hsindent{2}{}\<[9]%
\>[9]{}\neg \;(\Varid{same}\;[\mskip1.5mu \Varid{x}_{0},\Varid{x}_{1},\Varid{x}_{2}\mskip1.5mu]){}\<[E]%
\\
\>[7]{}\hsindent{2}{}\<[9]%
\>[9]{}\mathrel{\vee}\Varid{same}\;[\mskip1.5mu \Varid{x}_{3},\Varid{x}_{4}\mskip1.5mu]{}\<[E]%
\ColumnHook
\end{hscode}\resethooks
  \end{minipage}
};
\end{tikzpicture}
        \caption{The four polynomials computed by \ensuremath{\Varid{genAlgThinMemo}\;\mathrm{5}\;\Varid{sim}_{5}}.}
        \label{fig:4polys}
\end{figure}
To give the flavour of the end result we start with two examples which will be explained in detail later:
the level-\ensuremath{\Varid{p}}-complexity of 2-level iterated majority \ensuremath{\Varid{maj}_{3}^2} and of a 5-bit function we call \ensuremath{\Varid{sim}_{5}}, defined in \cref{fig:4polys}.
%
The level-$p$-complexity is a piecewise polynomial function of the probability \ensuremath{\Varid{p}} and \ensuremath{\Varid{sim}_{5}} is the smallest arity Boolean function we have found which has more than one polynomial piece contributing to the complexity.
Polynomials are represented by their coefficients: for example, \ensuremath{\Conid{P}\;[\mskip1.5mu \mathrm{5},\mathbin{-}\mathrm{8},\mathrm{8}\mskip1.5mu]} represents \(5-8x+8x^2\).
The function \ensuremath{\Varid{genAlgThinMemo}} uses thinning and memoization to generate a set of minimal cost polynomials.
\begin{hscode}\SaveRestoreHook
\column{B}{@{}>{\hspre}l<{\hspost}@{}}%
\column{11}{@{}>{\hspre}l<{\hspost}@{}}%
\column{25}{@{}>{\hspre}l<{\hspost}@{}}%
\column{30}{@{}>{\hspre}l<{\hspost}@{}}%
\column{31}{@{}>{\hspre}l<{\hspost}@{}}%
\column{49}{@{}>{\hspre}l<{\hspost}@{}}%
\column{67}{@{}>{\hspre}c<{\hspost}@{}}%
\column{67E}{@{}l@{}}%
\column{E}{@{}>{\hspre}l<{\hspost}@{}}%
\>[B]{}\Varid{ps5}\mathrel{=}\Varid{genAlgThinMemo}\;\mathrm{5}\;{}\<[25]%
\>[25]{}\Varid{sim}_{5}{}\<[31]%
\>[31]{}\mathbin{::}\Conid{Set}\;(\Conid{Poly}\;\mathbb{Q}){}\<[E]%
\\
\>[B]{}\Varid{check5}\mathrel{=}{}\<[11]%
\>[11]{}\Varid{ps5}\doubleequals\Varid{fromList}\;[\mskip1.5mu {}\<[30]%
\>[30]{}\Conid{P}\;[\mskip1.5mu \mathrm{2},\mathrm{6},\mathbin{-}\mathrm{10},\mathrm{8},\mathbin{-}\mathrm{4}\mskip1.5mu],{}\<[49]%
\>[49]{}\Conid{P}\;[\mskip1.5mu \mathrm{4},\mathbin{-}\mathrm{2},\mathbin{-}\mathrm{3},\mathrm{8},\mathbin{-}\mathrm{2}\mskip1.5mu],{}\<[E]%
\\
\>[30]{}\Conid{P}\;[\mskip1.5mu \mathrm{5},\mathbin{-}\mathrm{8},\mathrm{9},\mathrm{0},\mathbin{-}\mathrm{2}\mskip1.5mu],{}\<[49]%
\>[49]{}\Conid{P}\;[\mskip1.5mu \mathrm{5},\mathbin{-}\mathrm{8},\mathrm{8}\mskip1.5mu]{}\<[67]%
\>[67]{}\mskip1.5mu]{}\<[67E]%
\ColumnHook
\end{hscode}\resethooks
The graph, in \cref{fig:4polys}, shows that different polynomials dominate in different intervals.
The polynomial \ensuremath{\Conid{P}_{1}} is best near the end-points, but \ensuremath{\Conid{P}_{4}} is best near \ensuremath{\Varid{p}\mathrel{=}\nicefrac{1}{2}} (despite being really bad near the endpoints).
The level-\ensuremath{\Varid{p}}-complexity is the piecewise polynomial minimum, a combination of \ensuremath{\Conid{P}_{1}} and \ensuremath{\Conid{P}_{4}}.
This computation can be done by exhaustive search over the \ensuremath{\mathrm{54192}} different decision trees and \ensuremath{\mathrm{39}} resulting polynomials, but for more complex Boolean functions the doubly exponential growth makes that impractical.

For our running example, \ensuremath{\Varid{maj}_{3}^2}, a crude estimate indicates we would have $10^{111}$ decision trees to search and very many polynomials.
Thus the computation would be intractable if it were not for the combination of thinning, memoization, and symbolic comparison of polynomials.
Thanks to symmetries in the problem there turns out to be just one dominating polynomial:
\begin{hscode}\SaveRestoreHook
\column{B}{@{}>{\hspre}l<{\hspost}@{}}%
\column{11}{@{}>{\hspre}l<{\hspost}@{}}%
\column{25}{@{}>{\hspre}l<{\hspost}@{}}%
\column{30}{@{}>{\hspre}l<{\hspost}@{}}%
\column{31}{@{}>{\hspre}l<{\hspost}@{}}%
\column{60}{@{}>{\hspre}c<{\hspost}@{}}%
\column{60E}{@{}l@{}}%
\column{E}{@{}>{\hspre}l<{\hspost}@{}}%
\>[B]{}\Varid{ps9}\mathrel{=}\Varid{genAlgThinMemo}\;\mathrm{9}\;{}\<[25]%
\>[25]{}\Varid{maj}_{3}^2{}\<[31]%
\>[31]{}\mathbin{::}\Conid{Set}\;(\Conid{Poly}\;\mathbb{Q}){}\<[E]%
\\
\>[B]{}\Varid{check9}\mathrel{=}{}\<[11]%
\>[11]{}\Varid{ps9}\doubleequals\Varid{fromList}\;[\mskip1.5mu {}\<[30]%
\>[30]{}\Conid{P}\;[\mskip1.5mu \mathrm{4},\mathrm{4},\mathrm{6},\mathrm{9},\mathbin{-}\mathrm{61},\mathrm{23},\mathrm{67},\mathbin{-}\mathrm{64},\mathrm{16}\mskip1.5mu]{}\<[60]%
\>[60]{}\mskip1.5mu]{}\<[60E]%
\ColumnHook
\end{hscode}\resethooks
The graph, shown later in \cref{fig:itermajalgs2}, shows that only 4 bits are needed in the limiting cases of \ensuremath{\Varid{p}\mathrel{=}\mathrm{0}} or \ensuremath{\mathrm{1}} and that just over 6 bits are needed in the maximum at \ensuremath{\Varid{p}\mathrel{=}\nicefrac{1}{2}}.


\section{Background}
\label{sec:Background}
To explain what level-\ensuremath{\Varid{p}}-complexity of Boolean functions means we introduce some background about Boolean functions, decision trees, cost and complexity.
The Boolean input type \ensuremath{\mathbb{B}} could be \ensuremath{\{\mskip1.5mu \Conid{False},\Conid{True}\mskip1.5mu\},\{\mskip1.5mu \Conid{F},\Conid{T}\mskip1.5mu\}} or \ensuremath{\{\mskip1.5mu \mathrm{0},\mathrm{1}\mskip1.5mu\}} and from now on we use \F{} for false and \T{} for true in our notation.
In the running text we write \ensuremath{\Varid{e}\mathop{:}\Varid{t}} for ``\ensuremath{\Varid{e}} has type \ensuremath{\Varid{t}}'' which in the quoted Haskell code is written \ensuremath{\Varid{e}\mathbin{::}\Varid{t}}.

\subsection{Boolean functions}\label{sec:BoFun}
A Boolean function \ensuremath{\Varid{f}\mathop{:}\mathbb{B}^{\Varid{n}}\,\to\,\mathbb{B}} is a function from \ensuremath{\Varid{n}} Boolean inputs to one Boolean output.
We sometimes write \ensuremath{\Conid{BoolFun}\;\Varid{n}} for the type \ensuremath{\mathbb{B}^{\Varid{n}}\,\to\,\mathbb{B}}.
The easiest examples of Boolean functions are the functions \ensuremath{\ensuremath{\Varid{const}_{\Varid{n}}}\;\Varid{b}} which ignore the \ensuremath{\Varid{n}} input bits and return \ensuremath{\Varid{b}}.
The usual logical gates like \ensuremath{\Varid{and}_{\Varid{n}}} and \ensuremath{\Varid{or}_{\Varid{n}}} are very common Boolean functions.
Another example is the dictator function (also known as first projection), which is defined as \ensuremath{\ensuremath{\Varid{dict}_{\Varid{n}\mathbin{+}\mathrm{1}}}\;[\mskip1.5mu \Varid{x}_{0},\mathbin{...},\Varid{x}_n\mskip1.5mu]\mathrel{=}\Varid{x}_{0}} when the dictator is bit~\ensuremath{\mathrm{0}}.

A naive representation of a Boolean function could be a pair of an arity and a function \ensuremath{\Varid{f}} : \ensuremath{[\mskip1.5mu \mathbb{B}\mskip1.5mu]\,\to\,\mathbb{B}}, but that turns out to be inefficient when we want to compare and tabulate them (see \cref{sec:memo}).
Instead we use Binary Decision Diagrams \ensuremath{\Conid{BDD}}s \citep{Bryant_BDD_1986} as implemented in Masahiro Sakai's excellent Hackage package\footnote{\url{https://github.com/msakai/haskell-decision-diagrams}}.
The package reimplements all the usual Boolean operations on what is semantically expressions in \ensuremath{\Varid{n}} Boolean variables.
BDDs are an efficient way of representing Boolean functions, and they can be used for testing, verification and complexity analysis.
For readability, we will present Boolean functions in the naive represention, but the actual code uses the type \ensuremath{\Conid{BDD}\;\Varid{a}} from the \ensuremath{\Conid{BDD}} package (where \ensuremath{\Varid{a}} keeps track of variable ordering).
Note that we only use BDDs to represent our Boolean functions, not our decision trees.

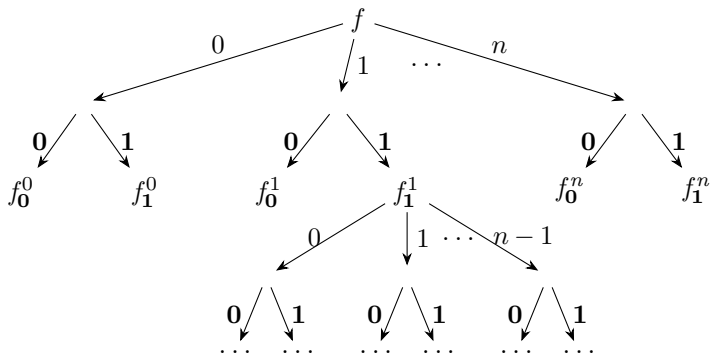
\begin{figure}[tbp]
  \centering
  \begin{forest}
    for tree = {
        minimum height=1ex, text depth = 0.25ex,
       anchor = north, 
         edge = {-Stealth},
        s sep = 3em,
        l sep = 2em
                },
    EL/.style = {
       before typesetting nodes={
    where n=1{edge label/.wrap value={node[pos=0.5,anchor=east]{#1}}}
             {edge label/.wrap value={node[pos=0.5,anchor=west]{#1}}}
                                }
                }
    [\ensuremath{\Varid{f}}, 
        [, edge label = {node[midway,anchor = south] {0}}
            [\ensuremath{\fixatill{\mathrm{0}}{\F{}}{\Varid{f}}}, EL = \ensuremath{\F{}}] 
            [\ensuremath{\fixatill{\mathrm{0}}{\T{}}{\Varid{f}}}, EL = \ensuremath{\T{}}] 
        ]
        [, EL = 1 
            [\ensuremath{\fixatill{\mathrm{1}}{\F{}}{\Varid{f}}}, EL = \ensuremath{\F{}}] 
            [\ensuremath{\fixatill{\mathrm{1}}{\T{}}{\Varid{f}}}, EL = \ensuremath{\T{}}, s sep = 1em
                [, EL = 0, s sep = 1ex
                [$\ldots$, EL = \ensuremath{\F{}}]
                    [$\ldots$, EL = \ensuremath{\T{}}]
                ]
                [, EL = 1, s sep = 1ex
                [$\ldots$, EL = \ensuremath{\F{}}]
                    [$\ldots$, EL = \ensuremath{\T{}}]
                ]
                [,no edge,EL = $\ldots$]
                [,no edge, EL = {}]
                [, EL = $n-1$, s sep = 1ex
                    [$\ldots$, EL = \ensuremath{\F{}}]
                    [$\ldots$, EL = \ensuremath{\T{}}]
                ]
            ] 
        ]
        [,no edge,EL = $\ldots$]
        [,no edge]
        [, edge label = {node[midway,anchor = south]{$n$}}
            [\ensuremath{\fixatill{\Varid{n}}{\F{}}{\Varid{f}}}, EL = \ensuremath{\F{}}] 
            [\ensuremath{\fixatill{\Varid{n}}{\T{}}{\Varid{f}}}, EL = \ensuremath{\T{}}] 
      ] 
    ]
\end{forest}
  \caption{The tree of subfunctions of a Boolean function \ensuremath{\Varid{f}}.
    This tree structure is also the call-graph for our generation of decision trees.
    Note that this tree structure is related to, but not the same as, the decision trees.}
  \label{fig:subf}
\end{figure}

%
%
In the complexity computation, we only need two operations on Boolean functions which we capture in the following type class interface:
\begin{hscode}\SaveRestoreHook
\column{B}{@{}>{\hspre}l<{\hspost}@{}}%
\column{3}{@{}>{\hspre}l<{\hspost}@{}}%
\column{12}{@{}>{\hspre}l<{\hspost}@{}}%
\column{E}{@{}>{\hspre}l<{\hspost}@{}}%
\>[B]{}\mathbf{class}\;\Conid{BoFun}\;\Varid{bf}\;\mathbf{where}{}\<[E]%
\\
\>[B]{}\hsindent{3}{}\<[3]%
\>[3]{}\Varid{isConst}{}\<[12]%
\>[12]{}\mathbin{::}\Varid{bf}\,\to\,\Conid{Maybe}\;\mathbb{B}{}\<[E]%
\\
\>[B]{}\hsindent{3}{}\<[3]%
\>[3]{}\Varid{setBit}{}\<[12]%
\>[12]{}\mathbin{::}\Conid{Index}\,\to\,\mathbb{B}\,\to\,\Varid{bf}\,\to\,\Varid{bf}{}\<[E]%
\\[\blanklineskip]%
\>[B]{}\mathbf{type}\;\Conid{Index}\mathrel{=}\mathbb{N}{}\<[E]%
\ColumnHook
\end{hscode}\resethooks
The use of a type class here means we keep the interface to the BDD implementation minimal, which makes proofs easier and gives better feedback from the type system.
The first method, \ensuremath{\Varid{isConst}\;\Varid{f}}, returns \ensuremath{\Conid{Just}\;\Varid{b}} iff the function \ensuremath{\Varid{f}} is constant and always returns \ensuremath{\Varid{b}\mathop{:}\mathbb{B}}.
The second method, \ensuremath{\Varid{setBit}\;\Varid{i}\;\Varid{b}\;\Varid{f}}, restricts a Boolean function (on \ensuremath{\Varid{n}\mathbin{+}\mathrm{1}} bits) by setting its \ensuremath{\Varid{i}}th bit to \ensuremath{\Varid{b}}.
The result is a ``subfunction'' on the remaining \ensuremath{\Varid{n}} bits, abbreviated $\fixatill{i}{b}{f}$, and illustrated in \cref{fig:subf}.

\begin{figure}[htbp]
  \centering
  \begin{forest}
    for tree = {
        minimum height=1ex, text depth = 0.25ex,
       anchor = north,
         edge = {-Stealth},
        s sep = 1em,
        l sep = 2em
                },
    EL/.style = {
       before typesetting nodes={
    where n=1{edge label/.wrap value={node[pos=0.5,anchor=east]{#1}}}
             {edge label/.wrap value={node[pos=0.5,anchor=west]{#1}}}
                                }
                }
    [\ensuremath{\ensuremath{\Varid{dict}_{\Varid{n}\mathbin{+}\mathrm{1}}}},
        [, edge label = {node[midway,anchor = south] {0}}
            [\ensuremath{\ensuremath{\Varid{const}_{\Varid{n}}}\;\F{}}, EL = \ensuremath{\F{}}]
            [\ensuremath{\ensuremath{\Varid{const}_{\Varid{n}}}\;\T{}}, EL = \ensuremath{\T{}}]
        ]
        [, EL = 1
            [\ensuremath{\ensuremath{\Varid{dict}_{\Varid{n}}}}, EL = \ensuremath{\F{}}]
            [\ensuremath{\ensuremath{\Varid{dict}_{\Varid{n}}}}, EL = \ensuremath{\T{}}]
        ]
        [,no edge,EL = $\ldots$]
        [,no edge]
        [, edge label = {node[midway,anchor = south]{$n$}}
            [\ensuremath{\ensuremath{\Varid{dict}_{\Varid{n}}}}, EL = \ensuremath{\F{}}]
            [\ensuremath{\ensuremath{\Varid{dict}_{\Varid{n}}}}, EL = \ensuremath{\T{}}]
      ]
    ]
\end{forest}
  \caption{The tree of subfunctions of the \ensuremath{\ensuremath{\Varid{dict}_{\Varid{n}\mathbin{+}\mathrm{1}}}} function.}
  \label{fig:dict}
\end{figure}
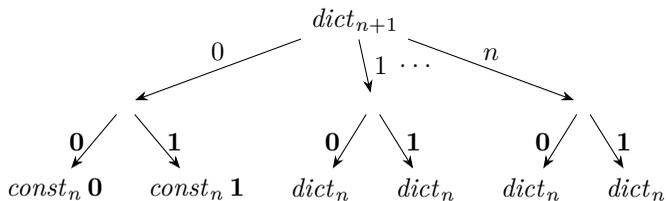

As an example, for the function \ensuremath{\Varid{and}_{2}} we have that \ensuremath{\Varid{setBit}\;\Varid{i}\;\F{}\;\Varid{and}_{2}\mathrel{=}\ensuremath{\Varid{const}_{\mathrm{1}}}\;\F{}} and \ensuremath{\Varid{setBit}\;\Varid{i}\;\T{}\;\Varid{and}_{2}\mathrel{=}\Varid{id}}.
For \ensuremath{\Varid{and}_{2}} we get the same result for \ensuremath{\Varid{i}\mathrel{=}\mathrm{0}}, or \ensuremath{\mathrm{1}} but for the
dictator function it depends if we pick the dictator index (\ensuremath{\mathrm{0}}) or
not.
We get \ensuremath{\Varid{setBit}\;\mathrm{0}\;\Varid{b}\;\ensuremath{\Varid{dict}_{\Varid{n}\mathbin{+}\mathrm{1}}}\mathrel{=}\ensuremath{\Varid{const}_{\Varid{n}}}\;\Varid{b}}, because the result is
dictated by bit \ensuremath{\mathrm{0}}.
Otherwise, we get \ensuremath{\Varid{setBit}\;(\Varid{i}\mathbin{+}\mathrm{1})\;\Varid{b}\;\ensuremath{\Varid{dict}_{\Varid{n}\mathbin{+}\mathrm{1}}}\mathrel{=}\ensuremath{\Varid{dict}_{\Varid{n}}}} irrespective
of the value of \ensuremath{\Varid{b}} since only the value of the dictator bit matters.
This behaviour is shown in \cref{fig:dict}.

\subsection{Decision trees}\label{sec:DecTree}
Consider a decision tree that picks the \ensuremath{\Varid{n}} bits of a Boolean function \ensuremath{\Varid{f}} in a deterministic way depending on the values of the bits picked further up the tree.
%
Decision trees are referred to as algorithms in \citep{landau2006lower,garban2014noise,jansson2022level}.
Given a Boolean function \ensuremath{\Varid{f}}, a decision tree \ensuremath{\Varid{t}} describes one way to evaluate the function \ensuremath{\Varid{f}}.
The Haskell datatype is as follows:
\begin{hscode}\SaveRestoreHook
\column{B}{@{}>{\hspre}l<{\hspost}@{}}%
\column{3}{@{}>{\hspre}l<{\hspost}@{}}%
\column{E}{@{}>{\hspre}l<{\hspost}@{}}%
\>[B]{}\mathbf{data}\;\Conid{DecTree}\mathrel{=}\Conid{Res}\;\mathbb{B}\mid \Conid{Pick}\;\Conid{Index}\;\Conid{DecTree}\;\Conid{DecTree}{}\<[E]%
\\
\>[B]{}\hsindent{3}{}\<[3]%
\>[3]{}\mathbf{deriving}\;(\Conid{Eq},\Conid{Ord},\Conid{Show}){}\<[E]%
\ColumnHook
\end{hscode}\resethooks
Parts of the ``rules of the game'' in the mathematical literature is that you must return a \ensuremath{\Conid{Res}}ult if the function is constant and you may only \ensuremath{\Conid{Pick}} an index once.
We can capture most of these rules with a type family version of the \ensuremath{\Conid{DecTree}} datatype (here expressed in \ensuremath{\Conid{Agda}} syntax).
Here we use two type indices: \ensuremath{\Varid{t}\mathop{:}\Conid{DecTree}\;\Varid{n}\;\Varid{f}} is a decision tree for the Boolean function \ensuremath{\Varid{f}}, of arity \ensuremath{\Varid{n}}.
The \ensuremath{\Conid{Res}} constructor may only be used for constant functions (but for any arity), while \ensuremath{\Conid{Pick}\;\Varid{i}} takes two subtrees for Boolean functions of arity \ensuremath{\Varid{n}} to a tree of arity \ensuremath{\Varid{suc}\;\Varid{n}\mathrel{=}\Varid{n}\mathbin{+}\mathrm{1}}.
\begin{hscode}\SaveRestoreHook
\column{B}{@{}>{\hspre}l<{\hspost}@{}}%
\column{3}{@{}>{\hspre}l<{\hspost}@{}}%
\column{9}{@{}>{\hspre}c<{\hspost}@{}}%
\column{9E}{@{}l@{}}%
\column{12}{@{}>{\hspre}l<{\hspost}@{}}%
\column{15}{@{}>{\hspre}c<{\hspost}@{}}%
\column{15E}{@{}l@{}}%
\column{18}{@{}>{\hspre}l<{\hspost}@{}}%
\column{40}{@{}>{\hspre}l<{\hspost}@{}}%
\column{62}{@{}>{\hspre}l<{\hspost}@{}}%
\column{88}{@{}>{\hspre}l<{\hspost}@{}}%
\column{E}{@{}>{\hspre}l<{\hspost}@{}}%
\>[B]{}\mathbf{data}\;\Conid{DecTree}{}\<[15]%
\>[15]{}\mathop{:}{}\<[15E]%
\>[18]{}(\Varid{n}\mathop{:}\mathbb{N})\,\to\,(\Varid{f}\mathop{:}\Conid{BoolFun}\;\Varid{n})\,\to\,\Conid{Set}\;\mathbf{where}{}\<[E]%
\\
\>[B]{}\hsindent{3}{}\<[3]%
\>[3]{}\Conid{Res}{}\<[9]%
\>[9]{}\mathop{:}{}\<[9E]%
\>[40]{}(\Varid{b}\mathop{:}\mathbb{B})\,\to\,{}\<[62]%
\>[62]{}\Conid{DecTree}\;\Varid{n}\;(\ensuremath{\Varid{const}_{\Varid{n}}}\;\Varid{b}){}\<[E]%
\\
\>[B]{}\hsindent{3}{}\<[3]%
\>[3]{}\Conid{Pick}{}\<[9]%
\>[9]{}\mathop{:}{}\<[9E]%
\>[12]{}\{\mskip1.5mu \Varid{f}\mathop{:}\Conid{BoolFun}\;(\Varid{suc}\;\Varid{n})\mskip1.5mu\}\,\to\,{}\<[40]%
\>[40]{}(\Varid{i}\mathop{:}\Conid{Fin}\;(\Varid{suc}\;\Varid{n}))\,\to\,{}\<[62]%
\>[62]{}\Conid{DecTree}\;\Varid{n}\;(\Varid{setBit}\;\Varid{i}\;\F{}\;\Varid{f})\,\to\,{}\<[E]%
\\
\>[62]{}\Conid{DecTree}\;\Varid{n}\;(\Varid{setBit}\;\Varid{i}\;\T{}\;{}\<[88]%
\>[88]{}\Varid{f})\,\to\,{}\<[E]%
\\
\>[62]{}\Conid{DecTree}\;(\Varid{suc}\;\Varid{n})\;\Varid{f}{}\<[E]%
\\[\blanklineskip]%
\>[B]{}\Varid{setBit}\mathop{:}\Conid{Fin}\;(\Varid{suc}\;\Varid{n})\,\to\,\mathbb{B}\,\to\,\Conid{BoolFun}\;(\Varid{suc}\;\Varid{n})\,\to\,\Conid{BoolFun}\;\Varid{n}{}\<[E]%
\ColumnHook
\end{hscode}\resethooks
Note that the dependently typed version of \ensuremath{\Varid{setBit}} clearly indicates that the resulting function \ensuremath{\Varid{g}\mathrel{=}(\Varid{setBit}\;\Varid{i}\;\Varid{b}\;\Varid{f})\mathop{:}\Conid{BoolFun}\;\Varid{n}} has arity one less that of \ensuremath{\Varid{f}\mathop{:}\Conid{BoolFun}\;(\Varid{suc}\;\Varid{n})}.
This helps maintaining the invariant that each input bit may only be picked once.%
\footnote{The use of \ensuremath{\Conid{Fin}\;\Varid{n}} also means that the interpretation of indices is local: the 3-bit example \ensuremath{\Varid{ex0}\mathrel{=}\Conid{Pick}\;\mathrm{0}\;(\Conid{Pick}\;\mathrm{0}\;(\Conid{Res}\;\F{}))\;(\Conid{Pick}\;\mathrm{1}\;(\Conid{Res}\;\T{}))} in Agda corresponds to the global interpretation \ensuremath{\Conid{Pick}\;\mathrm{0}\;(\Conid{Pick}\;\mathrm{1}\;(\Conid{Res}\;\F{}))\;(\Conid{Pick}\;\mathrm{2}\;(\Conid{Res}\;\T{}))}
. We use the global view in \ensuremath{\Varid{ex1}} and figures for readability.}
We use the Haskell versions, but the Agda versions capture the invariants better.

We can use these rules backwards to generate all possible decision trees for a certain function.
If the function is constant, returning \ensuremath{\Varid{b}\mathop{:}\mathbb{B}}, we immediately know that the only decision tree allowed is \ensuremath{\Conid{Res}\;\Varid{b}}.
If it is not constant, we pick any index \ensuremath{\Varid{i}}, any decision tree \ensuremath{\Varid{t}_{0}} for the subfunction \ensuremath{\Varid{setBit}\;\Varid{i}\;\F{}\;\Varid{f}} and \ensuremath{\Varid{t}_{1}} for the subfunction \ensuremath{\Varid{setBit}\;\Varid{i}\;\T{}\;\Varid{f}} recursively.
We get back to this in \cref{sec:genAlg} after some preparation.

Note that we do \textbf{not} use binary decision diagrams (BDDs) to
represent our decision trees.
\begin{figure}[tbp]
  \centering
\begin{forest}
    for tree = {
        draw, rounded corners,
        top color=white, bottom color=blue!20,
        font = \ttfamily,
        minimum height=1ex, text depth = 0.25ex,
       anchor = north, 
         edge = {-Stealth},
        s sep = 2em,
        l sep = 2em
                },
    EL/.style = {
       before typesetting nodes={
    where n=1{edge label/.wrap value={node[pos=0.5,anchor=east]{#1}}}
             {edge label/.wrap value={node[pos=0.5,anchor=west]{#1}}}
                                }
                }
    [$x_0$, root 
        [$x_2$, EL=\F
          [\F, EL=\F, sharp corners]
          [$x_1$, EL=\T
              [\F, EL=\F, sharp corners]
              [\T, EL=\T, sharp corners]]]
        [$x_1$, EL=\T 
            [$x_2$, EL=\F
              [\F, EL=\F, sharp corners]
              [\T, EL=\T, sharp corners]]
            [\T, EL=\T, sharp corners]]
    ]
\end{forest}


%
  \caption{An example of a decision tree for \ensuremath{\Varid{maj}_{3}}.
    The root node branches on the value of bit 0.
    If it is \F{}, it picks bit 2, while if it is \T{}, it picks bit 1.
    It then picks the last remaining bit if necessary.
  }
  \label{fig:Algex}
\end{figure}
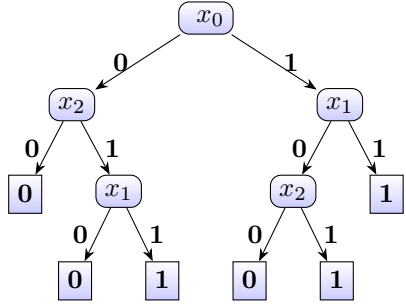
An example of a decision tree for the majority function \ensuremath{\Varid{maj}_{3}} on three bits
is defined by the expression \ensuremath{\Varid{ex1}} visualised in \cref{fig:Algex}.
\begin{hscode}\SaveRestoreHook
\column{B}{@{}>{\hspre}l<{\hspost}@{}}%
\column{15}{@{}>{\hspre}l<{\hspost}@{}}%
\column{36}{@{}>{\hspre}l<{\hspost}@{}}%
\column{56}{@{}>{\hspre}l<{\hspost}@{}}%
\column{E}{@{}>{\hspre}l<{\hspost}@{}}%
\>[B]{}\Varid{ex1}\mathrel{=}\Conid{Pick}\;\mathrm{0}\;{}\<[15]%
\>[15]{}(\Conid{Pick}\;\mathrm{2}\;(\Conid{Res}\;\F{})\;{}\<[36]%
\>[36]{}(\Conid{Pick}\;\mathrm{1}\;(\Conid{Res}\;\F{})\;(\Conid{Res}\;\T{})))\;{}\<[E]%
\\
\>[15]{}(\Conid{Pick}\;\mathrm{1}\;(\Conid{Pick}\;\mathrm{2}\;(\Conid{Res}\;\F{})\;(\Conid{Res}\;\T{}))\;{}\<[56]%
\>[56]{}(\Conid{Res}\;\T{})){}\<[E]%
\ColumnHook
\end{hscode}\resethooks

We will define several functions as folds over \ensuremath{\Conid{DecTree}} and to do that we introduce a type class \ensuremath{\Conid{TreeAlg}} (for ``Tree Algebra'') which collects the two methods \ensuremath{\Varid{res}} and \ensuremath{\Varid{pic}} which are then used in the fold to replace the constructors \ensuremath{\Conid{Res}} and \ensuremath{\Conid{Pick}}.
\begin{hscode}\SaveRestoreHook
\column{B}{@{}>{\hspre}l<{\hspost}@{}}%
\column{3}{@{}>{\hspre}l<{\hspost}@{}}%
\column{24}{@{}>{\hspre}l<{\hspost}@{}}%
\column{E}{@{}>{\hspre}l<{\hspost}@{}}%
\>[B]{}\mathbf{class}\;\Conid{TreeAlg}\;\Varid{a}\;\mathbf{where}{}\<[E]%
\\
\>[B]{}\hsindent{3}{}\<[3]%
\>[3]{}\Varid{res}\mathbin{::}\mathbb{B}\,\to\,\Varid{a}{}\<[E]%
\\
\>[B]{}\hsindent{3}{}\<[3]%
\>[3]{}\Varid{pic}\mathbin{::}\Conid{Index}\,\to\,\Varid{a}\,\to\,\Varid{a}\,\to\,\Varid{a}{}\<[E]%
\\[\blanklineskip]%
\>[B]{}\Varid{foldDT}\mathbin{::}\Conid{TreeAlg}\;\Varid{a}\Rightarrow \Conid{DecTree}\,\to\,\Varid{a}{}\<[E]%
\\
\>[B]{}\Varid{foldDT}\;(\Conid{Res}\;\Varid{b}){}\<[24]%
\>[24]{}\mathrel{=}\Varid{res}\;\Varid{b}{}\<[E]%
\\
\>[B]{}\Varid{foldDT}\;(\Conid{Pick}\;\Varid{i}\;\Varid{t}_{0}\;\Varid{t}_{1}){}\<[24]%
\>[24]{}\mathrel{=}\Varid{pic}\;\Varid{i}\;(\Varid{foldDT}\;\Varid{t}_{0})\;(\Varid{foldDT}\;\Varid{t}_{1}){}\<[E]%
\ColumnHook
\end{hscode}\resethooks
The \ensuremath{\Conid{TreeAlg}} class is used to define our decision trees but also for several other purposes.
(In the implementation we additionally require some total order on \ensuremath{\Varid{a}} to enable efficient set computations.)
We see that our decision tree type is the initial algebra of \ensuremath{\Conid{TreeAlg}} and that we can reimplement a generic version of \ensuremath{\Varid{ex1}} which can be instantiated to any \ensuremath{\Conid{TreeAlg}} instance:
\begin{hscode}\SaveRestoreHook
\column{B}{@{}>{\hspre}l<{\hspost}@{}}%
\column{14}{@{}>{\hspre}l<{\hspost}@{}}%
\column{34}{@{}>{\hspre}l<{\hspost}@{}}%
\column{53}{@{}>{\hspre}l<{\hspost}@{}}%
\column{E}{@{}>{\hspre}l<{\hspost}@{}}%
\>[B]{}\mathbf{instance}\;\Conid{TreeAlg}\;\Conid{DecTree}\;\mathbf{where}\;\Varid{res}\mathrel{=}\Conid{Res};\Varid{pic}\mathrel{=}\Conid{Pick};{}\<[E]%
\\[\blanklineskip]%
\>[B]{}\Varid{ex1}\mathbin{::}\Conid{TreeAlg}\;\Varid{a}\Rightarrow \Varid{a}{}\<[E]%
\\
\>[B]{}\Varid{ex1}\mathrel{=}\Varid{pic}\;\mathrm{0}\;{}\<[14]%
\>[14]{}(\Varid{pic}\;\mathrm{2}\;(\Varid{res}\;\F{})\;{}\<[34]%
\>[34]{}(\Varid{pic}\;\mathrm{1}\;(\Varid{res}\;\F{})\;(\Varid{res}\;\T{})))\;{}\<[E]%
\\
\>[14]{}(\Varid{pic}\;\mathrm{1}\;(\Varid{pic}\;\mathrm{2}\;(\Varid{res}\;\F{})\;(\Varid{res}\;\T{}))\;{}\<[53]%
\>[53]{}(\Varid{res}\;\T{})){}\<[E]%
\ColumnHook
\end{hscode}\resethooks

\subsection{Expected Cost}
For a function \ensuremath{\Varid{f}} and a specific input \ensuremath{\Varid{xs}\mathop{:}\mathbb{B}^{\Varid{n}}}, the cost of evaluating \ensuremath{\Varid{f}} according to a decision tree \ensuremath{\Varid{t}} is the length of the path from root to leaf dictated by the bits in \ensuremath{\Varid{xs}}.
We then let the bits be independent and identically distributed with probability $p \in [0,1]$ for \ensuremath{\T{}} and compute the \emph{expected} cost (averaging over all \(2^n\) inputs).
Expected cost can be implemented as an instance of \ensuremath{\Conid{TreeAlg}}.

\begin{joincode}%
\begin{hscode}\SaveRestoreHook
\column{B}{@{}>{\hspre}l<{\hspost}@{}}%
\column{E}{@{}>{\hspre}l<{\hspost}@{}}%
\>[B]{}\mathbf{newtype}\;\Conid{Poly}\;\Varid{a}\mathrel{=}\Conid{P}\;[\mskip1.5mu \Varid{a}\mskip1.5mu]{}\<[E]%
\ColumnHook
\end{hscode}\resethooks
\begin{hscode}\SaveRestoreHook
\column{B}{@{}>{\hspre}l<{\hspost}@{}}%
\column{62}{@{}>{\hspre}l<{\hspost}@{}}%
\column{E}{@{}>{\hspre}l<{\hspost}@{}}%
\>[B]{}\mathbf{type}\;\Conid{ExpCost}\;\Varid{a}\mathrel{=}\Conid{Poly}\;\Varid{a}{}\<[E]%
\\
\>[B]{}\mathbf{instance}\;\Conid{Ring}\;\Varid{a}\Rightarrow \Conid{TreeAlg}\;(\Conid{ExpCost}\;\Varid{a})\;\mathbf{where}\;\Varid{res}\mathrel{=}\Varid{resPoly};{}\<[62]%
\>[62]{}\Varid{pic}\mathrel{=}\Varid{pickPoly}{}\<[E]%
\\[\blanklineskip]%
\>[B]{}\Varid{expCost}\mathbin{::}\Conid{Ring}\;\Varid{a}\Rightarrow \Conid{DecTree}\,\to\,\Conid{Poly}\;\Varid{a}{}\<[E]%
\\
\>[B]{}\Varid{expCost}\mathrel{=}\Varid{foldDT}{}\<[E]%
\ColumnHook
\end{hscode}\resethooks
\end{joincode}
Note that the expected cost of any decision tree for a Boolean function of \ensuremath{\Varid{n}} bits will always be a polynomial.
We represent polynomials as lists of coefficients: \ensuremath{\Conid{P}\;[\mskip1.5mu \mathrm{1},\mathrm{2},\mathrm{3}\mskip1.5mu]} represents \ensuremath{\lambda \Varid{p}\,\to\,\mathrm{1}\mathbin{+}\mathrm{2}\times\Varid{p}\mathbin{+}\mathrm{3}\times\ensuremath{\Varid{p}^{\mathrm{2}}}}, and use \ensuremath{\Varid{evalP}\mathop{:}\Conid{Ring}\;\Varid{a}\Rightarrow \Conid{Poly}\;\Varid{a}\,\to\,(\Varid{a}\,\to\,\Varid{a})} to evaluate polynomials. The polynomial implementation borrowed from \cite{JanssonIonescuBernardyDSLsofMathBook2022} includes the polynomial ring operations (\ensuremath{(\mathbin{+})}, \ensuremath{(\mathbin{-})}, \ensuremath{(\times)}), \ensuremath{\Varid{gcd}}, \ensuremath{\Varid{divMod}}, symbolic derivative, and ordering.
The \ensuremath{\Varid{res}} and \ensuremath{\Varid{pic}} functions are as follows:
\begin{hscode}\SaveRestoreHook
\column{B}{@{}>{\hspre}l<{\hspost}@{}}%
\column{E}{@{}>{\hspre}l<{\hspost}@{}}%
\>[B]{}\Varid{resPoly}\mathbin{::}\Conid{Ring}\;\Varid{a}\Rightarrow \mathbb{B}\,\to\,\Varid{a}{}\<[E]%
\\
\>[B]{}\Varid{resPoly}\;\Varid{b}\mathrel{=}\Varid{zero}{}\<[E]%
\\[\blanklineskip]%
\>[B]{}\Varid{pickPoly}\mathbin{::}\Conid{Ring}\;\Varid{a}\Rightarrow \Conid{Index}\,\to\,\Conid{Poly}\;\Varid{a}\,\to\,\Conid{Poly}\;\Varid{a}\,\to\,\Conid{Poly}\;\Varid{a}{}\<[E]%
\\
\>[B]{}\Varid{pickPoly}\;\Varid{i}\;\Varid{q}_{0}\;\Varid{q}_{1}\mathrel{=}\Varid{one}\mathbin{+}(\Varid{one}\mathbin{-}\Varid{xP})\times\Varid{q}_{0}\mathbin{+}\Varid{xP}\times\Varid{q}_{1}{}\<[E]%
\ColumnHook
\end{hscode}\resethooks
Here \ensuremath{\Varid{zero}\mathrel{=}\Conid{P}\;[\mskip1.5mu \mskip1.5mu]} and \ensuremath{\Varid{one}\mathrel{=}\Conid{P}\;[\mskip1.5mu \mathrm{1}\mskip1.5mu]} represent \ensuremath{\Varid{const}\;\mathrm{0}} and \ensuremath{\Varid{const}\;\mathrm{1}} respectively while \ensuremath{\Varid{xP}\mathrel{=}\Conid{P}\;[\mskip1.5mu \mathrm{0},\mathrm{1}\mskip1.5mu]} is ``the polynomial \ensuremath{\Varid{x}}''.
For \ensuremath{\Varid{pickPoly}\;\anonymous \;\Varid{q}_{0}\;\Varid{q}_{1}} we first have to pick one bit and then if this bit is \ensuremath{\F{}} (with probability $\mathbb{P}(\val{i} = \F) = (1-p)$) we get $q_0$ which is the polynomial for this case.
If the bit is instead \ensuremath{\T{}} (with probability $\mathbb{P}(\val{i} = \T) = p$) we get $q_1$.
The expected cost of the decision tree \ensuremath{\Varid{ex1}} is $2  + 2p - 2p^2$.
From now on we will use Haskell's overloading to write \ensuremath{\mathrm{0}} and \ensuremath{\mathrm{1}} for \ensuremath{\Varid{zero}} and \ensuremath{\Varid{one}} even when working with polynomials.

\subsection{Complexity}\label{sec:complexity}
Now that we have introduced expected cost, we can introduce the level-\ensuremath{\Varid{p}}-complexity \ensuremath{\ensuremath{D_p(\Varid{f})}} as the pointwise minimum of the expected cost over all of \ensuremath{\Varid{f}}'s decision trees:
\begin{hscode}\SaveRestoreHook
\column{B}{@{}>{\hspre}l<{\hspost}@{}}%
\column{11}{@{}>{\hspre}l<{\hspost}@{}}%
\column{E}{@{}>{\hspre}l<{\hspost}@{}}%
\>[B]{}\ensuremath{D_p(\Varid{f})}\mathrel{=}\Varid{minimum}\;\{\mskip1.5mu \Varid{evalP}\;(\Varid{expCost}\;\Varid{t})\;\Varid{p}\mid \Varid{t}\leftarrow \ensuremath{\Varid{genAlg}_{\Varid{n}}}\;\Varid{f}\mskip1.5mu\}{}\<[E]%
\\[\blanklineskip]%
\>[B]{}\ensuremath{\Varid{genAlg}_{\Varid{n}}}{}\<[11]%
\>[11]{}\mathbin{::}(\Conid{BoFun}\;\Varid{bf},\Conid{TreeAlg}\;\Varid{a},\Conid{Ord}\;\Varid{a})\Rightarrow \Varid{bf}\,\to\,\Conid{Set}\;\Varid{a}{}\<[E]%
\ColumnHook
\end{hscode}\resethooks
where the generation of decision trees is explained in \cref{sec:genAlg}.
When minimizing we do not necessarily get a polynomial, but a piecewise polynomial function.
%
%
For simplicity we represent a piecewise polynomial function as a set of polynomials:
\begin{hscode}\SaveRestoreHook
\column{B}{@{}>{\hspre}l<{\hspost}@{}}%
\column{E}{@{}>{\hspre}l<{\hspost}@{}}%
\>[B]{}\mathbf{type}\;\Conid{PPoly}\;\Varid{a}\mathrel{=}\Conid{Set}\;(\Conid{Poly}\;\Varid{a}){}\<[E]%
\\
\>[B]{}\Varid{evalPP}\mathbin{::}(\Conid{Ring}\;\Varid{a},\Conid{Ord}\;\Varid{a})\Rightarrow \Conid{PPoly}\;\Varid{a}\,\to\,(\Varid{a}\,\to\,\Varid{a}){}\<[E]%
\\
\>[B]{}\Varid{evalPP}\;\Varid{qs}\;\Varid{p}\mathrel{=}\Varid{minimum}\;(\Varid{map}\;(\lambda \Varid{q}\,\to\,\Varid{evalP}\;\Varid{q}\;\Varid{p})\;\Varid{qs}){}\<[E]%
\ColumnHook
\end{hscode}\resethooks
This representation will be inefficient if the set is big, but as a specification it works fine and we will later use thinning to keep the set small (see \cref{sec:thin,sec:cmp}).
We say that one polynomial \ensuremath{\Varid{q}} is ``uniformly worse'' than another polynomial \ensuremath{\Varid{p}} when \ensuremath{\Varid{p}\;\Varid{x}\leq \Varid{q}\;\Varid{x}} for all \ensuremath{\mathrm{0}\leq \Varid{x}\leq \mathrm{1}} and \ensuremath{\Varid{p}\;\Varid{x}\mathbin{<}\Varid{q}\;\Varid{x}} for some \ensuremath{\mathrm{0}\mathbin{<}\Varid{x}\mathbin{<}\mathrm{1}}.
For some polynomials, we can not determine which is worse, see \cref{fig:4polys} where four polynomials all intersect.
In this case, they are incomparable.
%
%

When computing the level-\ensuremath{\Varid{p}}-complexity it would be possible to take both \ensuremath{\Varid{f}} and the probability \ensuremath{\Varid{p}} as arguments and return the smallest expected cost for that probability, but we prefer to just take \ensuremath{\Varid{f}} as an argument and compute a piecewise polynomial function representation.
In this way we can analyse the result symbolically to find minima, maxima, number of polynomial pieces, etc.

\subsection{Examples of Boolean functions and their costs}
\label{sec:ex}
Now that we have introduced expected cost and level-\ensuremath{\Varid{p}}-complexity we give a few examples of Boolean functions and their costs to give a feeling of how the computations work.
The impatient reader can skip forward to \cref{sec:method}.
As mentioned earlier (in \cref{sec:BoFun}), we present the Boolean functions as Haskell functions for readability, but every example has a BDD counterpart.

For the constant functions (\ensuremath{\ensuremath{\Varid{const}_{\Varid{n}}}\;\Varid{b}}), there is just one legal decision tree \ensuremath{\Varid{t}\mathrel{=}\Conid{Res}\;\Varid{b}} and thus \ensuremath{\Varid{expCost}\;\Varid{t}\mathrel{=}\mathrm{0}} which gives $D_p(\ensuremath{\ensuremath{\Varid{const}_{\Varid{n}}}\;\Varid{b}}) = 0$.
For the dictator function, there are many decision trees, but as we can see in \cref{fig:dict}, picking bit 0 first is optimal and gets us to the constant case just covered.
Thus the optimal tree is \ensuremath{\Varid{optTree}\mathrel{=}\Conid{Pick}\;\mathrm{0}\;(\Conid{Res}\;\F{})\;(\Conid{Res}\;\T{})} and we can compute the expected cost as follows.
\begin{hscode}\SaveRestoreHook
\column{B}{@{}>{\hspre}l<{\hspost}@{}}%
\column{10}{@{}>{\hspre}l<{\hspost}@{}}%
\column{19}{@{}>{\hspre}l<{\hspost}@{}}%
\column{E}{@{}>{\hspre}l<{\hspost}@{}}%
\>[B]{}\Varid{expCost}\;{}\<[10]%
\>[10]{}\Varid{optTree}{}\<[19]%
\>[19]{}\mathrel{=}\mathrm{1}\mathbin{+}(\mathrm{1}\mathbin{-}\Varid{xP})\times\mathrm{0}\mathbin{+}\Varid{xP}\times\mathrm{0}\mathrel{=}\mathrm{1}\ .{}\<[E]%
\ColumnHook
\end{hscode}\resethooks
which gives $D_p(\ensuremath{\ensuremath{\Varid{dict}_{\Varid{n}}}}) = 1$.

\begin{figure}[tbp]
  \centering
  \begin{forest}
    for tree = {
        minimum height=1ex, text depth = 0.25ex,
       anchor = north,
         edge = {-Stealth},
        s sep = 1em,
        l sep = 2em
                },
    EL/.style = {
       before typesetting nodes={
    where n=1{edge label/.wrap value={node[pos=0.5,anchor=east]{#1}}}
             {edge label/.wrap value={node[pos=0.5,anchor=west]{#1}}}
                                }
                }
    [\ensuremath{\Varid{par}_{\!\Varid{n}\mathbin{+}\mathrm{1}}},
        [, edge label = {node[midway,anchor = south] {0}}
            [\ensuremath{\Varid{par}_{\!\Varid{n}}}, EL = \ensuremath{\F{}}]
            [\ensuremath{\neg \;\Varid{par}_{\!\Varid{n}}}, EL = \ensuremath{\T{}}]
        ]
        [, EL = 1
            [\ensuremath{\Varid{par}_{\!\Varid{n}}}, EL = \ensuremath{\F{}}]
            [\ensuremath{\neg \;\Varid{par}_{\!\Varid{n}}}, EL = \ensuremath{\T{}}]
        ]
        [,no edge,EL = $\ldots$]
        [,no edge]
        [, edge label = {node[midway,anchor = south]{$n$}}
            [\ensuremath{\Varid{par}_{\!\Varid{n}}}, EL = \ensuremath{\F{}}]
            [\ensuremath{\neg \;\Varid{par}_{\!\Varid{n}}}, EL = \ensuremath{\T{}}]
      ]
    ]
\end{forest}
  \caption{The recursive structure of the parity function (\ensuremath{\Varid{par}_{\!\Varid{n}}}). The pattern repeats all the way down to \ensuremath{\Varid{par}_{\!\mathrm{0}}\mathrel{=}\ensuremath{\Varid{const}_{\mathrm{0}}}\;\F{}}.}
  \label{fig:par}
\end{figure}
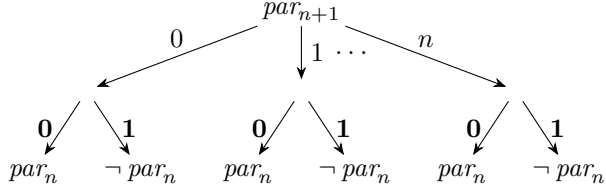

The parity function can be defined as
\begin{hscode}\SaveRestoreHook
\column{B}{@{}>{\hspre}l<{\hspost}@{}}%
\column{E}{@{}>{\hspre}l<{\hspost}@{}}%
\>[B]{}\Varid{count}\mathbin{::}\Conid{Eq}\;\Varid{a}\Rightarrow \Varid{a}\,\to\,[\mskip1.5mu \Varid{a}\mskip1.5mu]\,\to\,\Conid{Int}{}\<[E]%
\\
\>[B]{}\Varid{count}\;\Varid{x}\mathrel{=}\Varid{length}\mathbin{\circ}\Varid{filter}\;(\Varid{x}\doubleequals){}\<[E]%
\\[\blanklineskip]%
\>[B]{}\Varid{par}_{\!\Varid{n}}\mathbin{::}\mathbb{B}^{\Varid{n}}\,\to\,\mathbb{B}{}\<[E]%
\\
\>[B]{}\Varid{par}_{\!\Varid{n}}\mathrel{=}\Varid{odd}\mathbin{\circ}\Varid{count}\;\T{}{}\<[E]%
\ColumnHook
\end{hscode}\resethooks
In this case all bits have to be picked to determine the parity, regardless of input.
We prove that for all decision trees \ensuremath{\Varid{t}} of \ensuremath{\Varid{par}_{\!\Varid{n}}} or \ensuremath{\neg \;\Varid{par}_{\!\Varid{n}}} we have that \ensuremath{\Varid{expCost}\;\Varid{t}\mathrel{=}\Varid{n}} using induction over \ensuremath{\Varid{n}}. 
For the base case, \ensuremath{\Varid{n}\mathrel{=}\mathrm{0}} we have that \ensuremath{\Varid{par}_{\!\mathrm{0}}\mathrel{=}\ensuremath{\Varid{const}_{\mathrm{0}}}\;\F{}} and \ensuremath{\neg \;\Varid{par}_{\!\mathrm{0}}\mathrel{=}\ensuremath{\Varid{const}_{\mathrm{0}}}\;\T{}} so that \ensuremath{\Varid{expCost}\;\Varid{t}\mathrel{=}\mathrm{0}} for all decision trees \ensuremath{\Varid{t}} as shown above.
For the induction step we assume that for all decision trees \ensuremath{\Varid{t}} of \ensuremath{\Varid{par}_{\!\Varid{n}}} or \ensuremath{\neg \;\Varid{par}_{\!\Varid{n}}} we have that \ensuremath{\Varid{expCost}\;\Varid{t}\mathrel{=}\Varid{n}} and show that for all decision trees \ensuremath{\Varid{t}} of \ensuremath{\Varid{par}_{\!\Varid{n}\mathbin{+}\mathrm{1}}} or \ensuremath{\neg \;\Varid{par}_{\!\Varid{n}\mathbin{+}\mathrm{1}}} we have that \ensuremath{\Varid{expCost}\;\Varid{t}\mathrel{=}\Varid{n}\mathbin{+}\mathrm{1}}.
Any decision tree for \ensuremath{\Varid{par}_{\!\Varid{n}\mathbin{+}\mathrm{1}}} or \ensuremath{\neg \;\Varid{par}_{\!\Varid{n}\mathbin{+}\mathrm{1}}} is of the form \ensuremath{\Conid{Pick}\;\Varid{i}\;\Varid{t}_{0}\;\Varid{t}_{1}} where \ensuremath{\Varid{t}_{0}} and \ensuremath{\Varid{t}_{1}} are decision trees for \ensuremath{\Varid{par}_{\!\Varid{n}}} or \ensuremath{\neg \;\Varid{par}_{\!\Varid{n}}} as seen in \cref{fig:par}.
To calculate the expected cost we get
\begin{hscode}\SaveRestoreHook
\column{B}{@{}>{\hspre}l<{\hspost}@{}}%
\column{10}{@{}>{\hspre}l<{\hspost}@{}}%
\column{26}{@{}>{\hspre}l<{\hspost}@{}}%
\column{E}{@{}>{\hspre}l<{\hspost}@{}}%
\>[B]{}\Varid{expCost}\;{}\<[10]%
\>[10]{}(\Conid{Pick}\;\Varid{i}\;\Varid{t}_{0}\;\Varid{t}_{1}){}\<[26]%
\>[26]{}\mathrel{=}\mathrm{1}\mathbin{+}(\mathrm{1}\mathbin{-}\Varid{xP})\times(\Varid{expCost}\;\Varid{t}_{0})\mathbin{+}\Varid{xP}\times(\Varid{expCost}\;\Varid{t}_{1}){}\<[E]%
\\
\>[26]{}\mathrel{=}\mathrm{1}\mathbin{+}(\mathrm{1}\mathbin{-}\Varid{xP})\times\Varid{n}\mathbin{+}\Varid{xP}\times\Varid{n}\mathrel{=}\mathrm{1}\mathbin{+}\Varid{n}{}\<[E]%
\ColumnHook
\end{hscode}\resethooks
Thus, the induction proof is complete and as \ensuremath{\Varid{expCost}\;\Varid{t}\mathrel{=}\Varid{n}} for all decision trees then also the minimum is $n$, thus $D_p(\ensuremath{\Varid{par}_{\!\Varid{n}}}) = n$.
Comparing \cref{fig:dict} with \cref{fig:par}, we see that the minimum depth of the dictator tree is 1, while the minimum depth of the parity tree is $n$.
The parity function and the constant function are interesting extreme cases of Boolean functions as they have highest and lowest possible level-\ensuremath{\Varid{p}}-complexity \ensuremath{\Varid{n}} and 0.
Either all bits have to be picked to determine the parity, or none of them need to be picked to determine the constant function.

We now introduce the Boolean function \ensuremath{\Varid{same}} which checks if all bits are equal:
\begin{hscode}\SaveRestoreHook
\column{B}{@{}>{\hspre}l<{\hspost}@{}}%
\column{E}{@{}>{\hspre}l<{\hspost}@{}}%
\>[B]{}\Varid{same}\mathbin{::}\mathbb{B}^{\Varid{n}}\,\to\,\mathbb{B}{}\<[E]%
\\
\>[B]{}\Varid{same}\;\Varid{bs}\mathrel{=}\Varid{and}\;\Varid{bs}\mathrel{\vee}\neg \;(\Varid{or}\;\Varid{bs}){}\<[E]%
\ColumnHook
\end{hscode}\resethooks
Using \ensuremath{\Varid{same}} we construct the example \ensuremath{\Varid{sim}_{5}} from the introduction.
We first split the bits into two groups, one with the first three bits and the second with the last two bits.
On the first group, called \ensuremath{\Varid{as}}, we check if the bits are not the same, and on the second group, called \ensuremath{\Varid{cs}} we check if the bits are the same.
\begin{hscode}\SaveRestoreHook
\column{B}{@{}>{\hspre}l<{\hspost}@{}}%
\column{3}{@{}>{\hspre}l<{\hspost}@{}}%
\column{E}{@{}>{\hspre}l<{\hspost}@{}}%
\>[B]{}\Varid{sim}_{5}\mathbin{::}\mathbb{B}^{\mathrm{5}}\,\to\,\mathbb{B}{}\<[E]%
\\
\>[B]{}\Varid{sim}_{5}\;\Varid{bs}\mathrel{=}\neg \;(\Varid{same}\;\Varid{as})\mathrel{\vee}\Varid{same}\;\Varid{cs}{}\<[E]%
\\
\>[B]{}\hsindent{3}{}\<[3]%
\>[3]{}\mathbf{where}\;(\Varid{as},\Varid{cs})\mathrel{=}\Varid{splitAt}\;\mathrm{3}\;\Varid{bs}{}\<[E]%
\ColumnHook
\end{hscode}\resethooks
%
The point of this function is to illustrate a special case where the best decision tree depends on \ensuremath{\Varid{p}} so that the level-\ensuremath{\Varid{p}}-complexity consists of more than one polynomial piece.
This computation is shown in \cref{sec:fAC}.

One of the major goals of this paper was to calculate the level-\ensuremath{\Varid{p}}-complexity of 9 bit iterated majority called \ensuremath{\Varid{maj}_{3}^2}.
When extending the majority function to \ensuremath{\Varid{maj}_{3}^2}, we use \ensuremath{\Varid{maj}_{3}} inside \ensuremath{\Varid{maj}_{3}}.
\begin{hscode}\SaveRestoreHook
\column{B}{@{}>{\hspre}l<{\hspost}@{}}%
\column{3}{@{}>{\hspre}l<{\hspost}@{}}%
\column{10}{@{}>{\hspre}l<{\hspost}@{}}%
\column{17}{@{}>{\hspre}l<{\hspost}@{}}%
\column{23}{@{}>{\hspre}l<{\hspost}@{}}%
\column{E}{@{}>{\hspre}l<{\hspost}@{}}%
\>[B]{}\Varid{maj}_{3}^2\mathbin{::}\mathbb{B}^{\mathrm{9}}\,\to\,\mathbb{B}{}\<[E]%
\\
\>[B]{}\Varid{maj}_{3}^2\;\Varid{bs}\mathrel{=}\Varid{maj}_{3}\;[\mskip1.5mu \Varid{maj}_{3}\;\Varid{bs}_{0},\Varid{maj}_{3}\;\Varid{bs}_{1},\Varid{maj}_{3}\;\Varid{bs}_{2}\mskip1.5mu]{}\<[E]%
\\
\>[B]{}\hsindent{3}{}\<[3]%
\>[3]{}\mathbf{where}\;{}\<[10]%
\>[10]{}(\Varid{bs}_{0},{}\<[17]%
\>[17]{}\Varid{rest}{}\<[23]%
\>[23]{})\mathrel{=}\Varid{splitAt}\;\mathrm{3}\;\Varid{bs}{}\<[E]%
\\
\>[10]{}(\Varid{bs}_{1},{}\<[17]%
\>[17]{}\Varid{bs}_{2}{}\<[23]%
\>[23]{})\mathrel{=}\Varid{splitAt}\;\mathrm{3}\;\Varid{rest}{}\<[E]%
\\[\blanklineskip]%
\>[B]{}\Varid{maj}_{\Varid{n}}\mathbin{::}\mathbb{B}^{\Varid{n}}\,\to\,\mathbb{B}{}\<[E]%
\\
\>[B]{}\Varid{maj}_{\Varid{n}}\;\Varid{bs}\mathrel{=}\Varid{count}\;\T{}\;\Varid{bs}\geq \Varid{count}\;\F{}\;\Varid{bs}{}\<[E]%
\ColumnHook
\end{hscode}\resethooks
It is hard to calculate $D_p(\ensuremath{\Varid{maj}_{3}^2})$ by hand because there are very many different decision trees, and this motivated our Haskell implementation.


\section{Computing the level-$p$-complexity}
\label{sec:method}
In this section we explain how to compute the level-\ensuremath{\Varid{p}}-complexity of a Boolean function \ensuremath{\Varid{f}} by recursively “generating all candidates” followed by “picking the best one(s)”.
The naive approach would be to generate all decision trees of \ensuremath{\Varid{f}} and then minimizing, but already for the 9-bit function \ensuremath{\Varid{maj}_{3}^2} that is intractable.
To reduce the number of polynomials we use the algorithm design technique thinning.
We compare polynomials by using Yun's algorithm and Descartes rule of signs.
Further, since the same subfunctions often appear in many different nodes we can save a significant amount of computation time using memoization.

The top level complexity computation (from \cref{sec:complexity}) can be simplified a bit:
\begin{hscode}\SaveRestoreHook
\column{B}{@{}>{\hspre}l<{\hspost}@{}}%
\column{8}{@{}>{\hspre}c<{\hspost}@{}}%
\column{8E}{@{}l@{}}%
\column{11}{@{}>{\hspre}l<{\hspost}@{}}%
\column{E}{@{}>{\hspre}l<{\hspost}@{}}%
\>[B]{}\ensuremath{D_p(\Varid{f})}{}\<[8]%
\>[8]{}\mathrel{=}{}\<[8E]%
\>[11]{}\Varid{minimum}\;\{\mskip1.5mu \Varid{evalP}\;(\Varid{expCost}\;\Varid{t})\;\Varid{p}\mid \Varid{t}\leftarrow \ensuremath{\Varid{genAlg}_{\Varid{n}}}\;\Varid{f}\mskip1.5mu\}{}\<[E]%
\\
\>[11]{}\mbox{\commentbegin  comprehension syntax  \commentend}{}\<[E]%
\\
\>[8]{}\mathrel{=}{}\<[8E]%
\>[11]{}\Varid{minimum}\;(\Varid{map}\;(\lambda \Varid{t}\,\to\,\Varid{evalP}\;(\Varid{expCost}\;\Varid{t})\;\Varid{p})\;(\ensuremath{\Varid{genAlg}_{\Varid{n}}}\;\Varid{f})){}\<[E]%
\\
\>[11]{}\mbox{\commentbegin  \ensuremath{\Varid{map}\;(\Varid{g}\mathbin{\circ}\Varid{h})\mathrel{=}\Varid{map}\;\Varid{g}\mathbin{\circ}\Varid{map}\;\Varid{h}}  \commentend}{}\<[E]%
\\
\>[8]{}\mathrel{=}{}\<[8E]%
\>[11]{}\Varid{minimum}\;(\Varid{map}\;(\lambda \Varid{q}\,\to\,\Varid{evalP}\;\Varid{q}\;\Varid{p})\;(\Varid{map}\;\Varid{expCost}\;(\ensuremath{\Varid{genAlg}_{\Varid{n}}}\;\Varid{f}))){}\<[E]%
\\
\>[11]{}\mbox{\commentbegin  fuse \ensuremath{\Varid{expCost}} into the tree algebra generation  \commentend}{}\<[E]%
\\
\>[8]{}\mathrel{=}{}\<[8E]%
\>[11]{}\Varid{minimum}\;(\Varid{map}\;(\lambda \Varid{q}\,\to\,\Varid{evalP}\;\Varid{q}\;\Varid{p})\;(\ensuremath{\Varid{genAlg}_{\Varid{n}}}\;\Varid{f})){}\<[E]%
\\
\>[11]{}\mbox{\commentbegin  let \ensuremath{\Varid{best}\;\Varid{p}\mathrel{=}\Varid{minimum}\mathbin{\circ}\Varid{map}\;(\lambda \Varid{q}\,\to\,\Varid{evalP}\;\Varid{q}\;\Varid{p})}  \commentend}{}\<[E]%
\\
\>[8]{}\mathrel{=}{}\<[8E]%
\>[11]{}\Varid{best}\;\Varid{p}\;(\ensuremath{\Varid{genAlg}_{\Varid{n}}}\;\Varid{f}){}\<[E]%
\ColumnHook
\end{hscode}\resethooks
and we start by explaining \ensuremath{\ensuremath{\Varid{genAlg}_{\Varid{n}}}}.
The decision trees of a function \ensuremath{\Varid{f}} can be described in terms of the decision trees for the immediate subfunctions ($\fixatill{i}{b}{f} = \ensuremath{\Varid{setBit}\;\Varid{i}\;\Varid{b}\;\Varid{f}}$) for different \ensuremath{\Varid{i}\mathop{:}\Conid{Index}} and \ensuremath{\Varid{b}\mathop{:}\mathbb{B}}.
In fact, we can immediately generate elements of any tree algebra, not only decision trees, by using \ensuremath{\Varid{res}} and \ensuremath{\Varid{pic}} instead of \ensuremath{\Conid{Res}} and \ensuremath{\Conid{Pick}}.
(That is used in the ``fuse'' step of the calculation above.)
When we explain the algorithm we write ``decision tree'' to make it feel more concrete, but we will in the end mostly use it to directly compute expected cost polynomials.

\subsection{Generating decision trees and other tree algebras}\label{sec:genAlg}

The complexity computation starts from a Boolean function \ensuremath{\Varid{f}\mathop{:}\Conid{BoolFun}\;\Varid{n}}, and generates all decision trees for it.
There are two top level cases: either the function \ensuremath{\Varid{f}} is constant (and returns \ensuremath{\Varid{b}\mathop{:}\mathbb{B}}), in which case there is only one decision tree: \ensuremath{\Varid{res}\;\Varid{b}};
or the function \ensuremath{\Varid{f}} still depends on some of the input bits (and thus the arity is at least 1).
In the latter case, for each index \ensuremath{\Varid{i}\mathop{:}\Conid{Index}} we can generate two subfunctions $\fixatill{i}{\F}{f} = \ensuremath{\Varid{setBit}\;\Varid{i}\;\F{}\;\Varid{f}}$ and $\fixatill{i}{\T}{f}$ = \ensuremath{\Varid{setBit}\;\Varid{i}\;\T{}\;\Varid{f}}.
We then recursively generate a decision tree \ensuremath{\Varid{t}_{0}} for $\fixatill{i}{\F}{f}$ and \ensuremath{\Varid{t}_{1}} for $\fixatill{i}{\T}{f}$ and combine them to a bigger decision tree using \ensuremath{\Varid{pic}\;\Varid{i}\;\Varid{t}_{0}\;\Varid{t}_{1}}.
This is done for all combinations of \ensuremath{\Varid{i}}, \ensuremath{\Varid{t}_{0}}, and \ensuremath{\Varid{t}_{1}} in a set comprehension.
To make it easier to later extend the definition (for thinning and memoization) we make the recursive step explicit.
\begin{hscode}\SaveRestoreHook
\column{B}{@{}>{\hspre}l<{\hspost}@{}}%
\column{13}{@{}>{\hspre}l<{\hspost}@{}}%
\column{19}{@{}>{\hspre}l<{\hspost}@{}}%
\column{26}{@{}>{\hspre}l<{\hspost}@{}}%
\column{68}{@{}>{\hspre}l<{\hspost}@{}}%
\column{E}{@{}>{\hspre}l<{\hspost}@{}}%
\>[B]{}\Varid{genAlg}\mathbin{::}(\Conid{BoFun}\;\Varid{bf},\Conid{TreeAlg}\;\Varid{a},\Conid{Ord}\;\Varid{a})\Rightarrow \mathbb{N}\,\to\,\Varid{bf}\,\to\,\Conid{Set}\;\Varid{a}{}\<[E]%
\\
\>[B]{}\Varid{genAlg}\mathrel{=}\Varid{genAlgStep}\;\Varid{genAlg}{}\<[E]%
\\[\blanklineskip]%
\>[B]{}\Varid{genAlgStep}\mathbin{::}(\Conid{BoFun}\;\Varid{bf},\Conid{TreeAlg}\;\Varid{a},\Conid{Ord}\;\Varid{a})\Rightarrow (\mathbb{N}\,\to\,\Varid{bf}\,\to\,\Conid{Set}\;\Varid{a})\,\to\,(\mathbb{N}\,\to\,\Varid{bf}\,\to\,\Conid{Set}\;\Varid{a}){}\<[E]%
\\
\>[B]{}\Varid{genAlgStep}\;{}\<[13]%
\>[13]{}\Varid{genA}\;{}\<[19]%
\>[19]{}\Varid{n}\;{}\<[26]%
\>[26]{}\Varid{f}\mid \Conid{Just}\;\Varid{b}\leftarrow \Varid{isConst}\;\Varid{f}\mathrel{=}\{\Varid{res}\;\Varid{b}\}{}\<[E]%
\\
\>[B]{}\Varid{genAlgStep}\;{}\<[13]%
\>[13]{}\Varid{genA}\;{}\<[19]%
\>[19]{}(\Varid{n}\mathbin{+}\mathrm{1})\;{}\<[26]%
\>[26]{}\Varid{f}\mathrel{=} \{\mskip1.5mu \Varid{pic}\;\Varid{i}\;\Varid{t}_{0}\;\Varid{t}_{1}\mid \Varid{i}\leftarrow \{\mskip1.5mu \mathrm{0}\mathinner{\ldotp\ldotp}\Varid{n}\mskip1.5mu\},{}\<[68]%
\>[68]{}\Varid{t}_{0}\leftarrow \Varid{genA}\;\Varid{n}\;\fixatill{\Varid{i}}{\F{}}{\Varid{f}},\Varid{t}_{1}\leftarrow \Varid{genA}\;\Varid{n}\;\fixatill{\Varid{i}}{\T{}}{\Varid{f}}\mskip1.5mu\}{}\<[E]%
\ColumnHook
\end{hscode}\resethooks
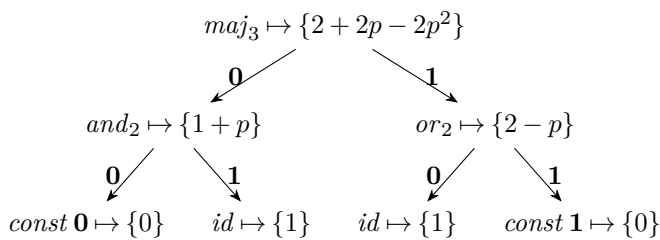
\begin{figure}[bp]
  \centering
  \begin{forest}
    for tree = {
       anchor = north,
         edge = {-Stealth},
        s sep = 1em,
        l sep = 2em
                },
    EL/.style = {
       before typesetting nodes={
    where n=1{edge label/.wrap value={node[pos=0.5,anchor=east]{#1}}}
             {edge label/.wrap value={node[pos=0.5,anchor=west]{#1}}}
                                }
                }
    [\genAlgNode{3}{\ensuremath{\Varid{maj}_{3}}}{\{2 + 2p - 2p^2\}},
        [\genAlgNode{2}{\ensuremath{\Varid{and}_{2}}}{\{1+ p\}}, EL = \ensuremath{\F{}}
            [\genAlgNode{1}{\ensuremath{\Varid{const}\;\F{}}}{\{0\}}, EL = \ensuremath{\F{}}]
            [\genAlgNode{1}{\ensuremath{\Varid{id}}}{\{1\}}, EL = \ensuremath{\T{}}
            ]
        ]
        [\genAlgNode{2}{\ensuremath{\Varid{or}_{2}}}{\{2 -p\}}, EL = \ensuremath{\T{}}
            [\genAlgNode{1}{\ensuremath{\Varid{id}}}{\{1\}}, EL = \ensuremath{\F{}}
            ]
            [\genAlgNode{1}{\ensuremath{\Varid{const}\;\T{}}}{\{0\}}, EL = \ensuremath{\T{}}]
        ]
    ]
\end{forest}

  \caption{A simplified computation tree of \ensuremath{\ensuremath{\Varid{genAlg}_{\mathrm{3}}}\;\Varid{maj}_{3}}. In each
    node \genAlgNode{}{\ensuremath{\Varid{f}}}{\ensuremath{\Varid{ps}}} shows the input \ensuremath{\Varid{f}} and output \ensuremath{\Varid{ps}\mathrel{=}\ensuremath{\Varid{genAlg}_{\Varid{n}}}\;\Varid{f}} of each local call. As all the functions involved are
    ``symmetric'' in the index (\ensuremath{\Varid{setBit}\;\Varid{i}\;\Varid{b}\;\Varid{f}\doubleequals\Varid{setBit}\;\Varid{j}\;\Varid{b}\;\Varid{f}} for all
    \ensuremath{\Varid{i}} and \ensuremath{\Varid{j}}) we only show edges for \ensuremath{\F{}} and \ensuremath{\T{}} from each
    level.}

  \label{fig:alg}
\end{figure}

We would like to enumerate the cost polynomials of all the decision trees of a particular Boolean function (\ensuremath{\Varid{n}\mathrel{=}\mathrm{9}}, \ensuremath{\Varid{f}\mathrel{=}\Varid{maj}_{3}^2} is our main goal).
Without taking symmetries into account there are \ensuremath{\mathrm{2}\times\Varid{n}} immediate subfunctions $\fixatill{i}{b}{f}$ and if $T_g$ is the cardinality of the enumeration for subfunction $g$ we have that
\nopagebreak
  $$T_{\ensuremath{\Varid{f}}} = \sum_{i=0}^{n-1} T_{\fixatill{i}{\F}{f}}\times T_{\fixatill{i}{\T}{f}}$$

These numbers can be really big if we count all decision trees, but if we only care about their cost polynomials, many decision trees will collapse to the same polynomial, making the counts more manageable (but still possibly really big).
Even the total number of subfunctions encountered (the number of recursive calls) can be quite big.
If all the \ensuremath{\mathrm{2}\times\Varid{n}} immediate subfunctions are different, and if all of them would generate \ensuremath{\mathrm{2}\times(\Varid{n}\mathbin{-}\mathrm{1})} different subfunctions in turn, the number of subfunctions would be $2^n \times n!$.
But in practice many subfunctions will be the same.
When computing the polynomials for the 9-bit function \ensuremath{\Varid{maj}_{3}^2}, for example, only \ensuremath{\mathrm{215}} distinct subfunctions are encountered.

As a smaller example, for the 3-bit majority function \ensuremath{\Varid{maj}_{3}}, choosing \ensuremath{\Varid{i}\mathrel{=}\mathrm{0},\mathrm{1},} or \ensuremath{\mathrm{2}} gives exactly the same subfunctions.
\Cref{fig:alg} illustrates a simplified call graph of \ensuremath{\ensuremath{\Varid{genAlg}_{\mathrm{3}}}\;\Varid{maj}_{3}} and the results (the expected cost polynomials) for the different subfunctions.
In this case all the sets are singletons, but that is very unusual for more realistic Boolean functions.
%
It would take too long to compute all polynomials for the 9-bit function \ensuremath{\Varid{maj}_{3}^2} but there are 21 distinct 7-bit sub-functions, and the first one of them already has \ensuremath{\mathrm{18021}} polynomials.
Thus we can expect billions of polynomials for \ensuremath{\Varid{maj}_{3}^2} and this means we need to look at ways to keep only the most promising candidates at each level.
This leads us to the algorithmic design technique of thinning.

\subsection{Thinning}\label{sec:thin}

The general shape of the specification has two phases: ``generate all candidates'' followed by ``pick the best one(s)''.
The first phase is recursive and we would like to push as much as possible of ``pick the best'' into the recursive computation.
In the extreme case of a greedy algorithm, we can thin the intermediate sets all the way down to singletons, but even if the sets are a bit bigger than that we can still reduce the computation cost significantly.
A good (but abstract) reference for thinning is the Algebra of Programming book \cite[Chapter 8]{bird_algebra_1997} and more concrete references are the corresponding developments in Agda \citep{DBLP:journals/jfp/MuKJ09} and Haskell \citep{bird_gibbons_2020}.
In this subsection the main focus is on specification and correctness, with Agda-like syntax for the logic part.


The ``pick the best'' phase is \ensuremath{\Varid{best}\;\Varid{p}\mathrel{=}\Varid{minimum}\mathbin{\circ}\Varid{map}\;(\lambda \Varid{q}\,\to\,\Varid{evalP}\;\Varid{q}\;\Varid{p})} of type \ensuremath{\Conid{Set}\;(\Conid{Poly}\;\Varid{r})\,\to\,\Varid{r}} for some ring of scalars \ensuremath{\Varid{r}} (usually rational numbers).
In this context it is clear that in the generation phase we can throw away any polynomial which is ``uniformly worse'' than some other polynomial and this is what we want to use thinning for.
We are looking for some ``smallest'' polynomials, but we only have a preorder, not a total order, which means that we may need to keep a set of incomparable candidates (elements \ensuremath{\Varid{x}\not\doubleequals\Varid{y}} for which neither \ensuremath{\Varid{x}\prec\Varid{y}} nor \ensuremath{\Varid{y}\prec\Varid{x}}).
We first describe the general setting and move to the specifics of our polynomials later.

We start from a strict preorder \ensuremath{(\prec)\mathop{:}\Varid{a}\,\to\,\Varid{a}\,\to\,\Conid{Prop}} (an irreflexive and transitive relation).
You can think of \ensuremath{\Conid{Prop}} as \ensuremath{\mathbb{B}} because we only work with decidable relations and finite sets in this application.
As we are looking for minima, we say that \ensuremath{\Varid{y}} \emph{dominates} \ensuremath{\Varid{x}} if \ensuremath{\Varid{y}\prec\Varid{x}}.
%

We lift the order relation to sets in two steps.
First \ensuremath{\Varid{ys}\mathrel{\dot{\prec}}\Varid{x}} means that \ensuremath{\Varid{ys}} \emph{dominates} \ensuremath{\Varid{x}},
meaning that some element in \ensuremath{\Varid{ys}} is smaller than \ensuremath{\Varid{x}}.
If this holds, there is no need to add \ensuremath{\Varid{x}} to \ensuremath{\Varid{ys}} because we already
have at least one better element in \ensuremath{\Varid{ys}}.
Then \ensuremath{\Varid{ys}\mathrel{\ddot{\prec}}\Varid{xs}} means that \ensuremath{\Varid{ys}} dominates all of \ensuremath{\Varid{xs}}.
\begin{hscode}\SaveRestoreHook
\column{B}{@{}>{\hspre}l<{\hspost}@{}}%
\column{E}{@{}>{\hspre}l<{\hspost}@{}}%
\>[B]{}(\mathrel{\dot{\prec}})\mathop{:}\Conid{Set}\;\Varid{a}\,\to\,\Varid{a}\,\to\,\Conid{Prop}{}\<[E]%
\\
\>[B]{}\Varid{ys}\mathrel{\dot{\prec}}\Varid{x}\mathrel{=}\exists\,\Varid{y}\in \Varid{ys}.\,\,\Varid{y}\prec\Varid{x}{}\<[E]%
\\[\blanklineskip]%
\>[B]{}(\mathrel{\ddot{\prec}})\mathop{:}\Conid{Set}\;\Varid{a}\,\to\,\Conid{Set}\;\Varid{a}\,\to\,\Conid{Prop}{}\<[E]%
\\
\>[B]{}\Varid{ys}\mathrel{\ddot{\prec}}\Varid{xs}\mathrel{=}\forall\,\Varid{x}\in \Varid{xs}.\,\,\Varid{ys}\mathrel{\dot{\prec}}\Varid{x}{}\<[E]%
\ColumnHook
\end{hscode}\resethooks
Finally, we combine subset and domination into the thinning relation:
\begin{hscode}\SaveRestoreHook
\column{B}{@{}>{\hspre}l<{\hspost}@{}}%
\column{E}{@{}>{\hspre}l<{\hspost}@{}}%
\>[B]{}\Conid{Thin}\;\Varid{ys}\;\Varid{xs}\mathrel{=}(\Varid{ys}\subseteq\Varid{xs})\mathrel{\wedge}\Varid{ys}\mathrel{\ddot{\prec}}(\Varid{xs}\mathbin{\char92 \char92 }\Varid{ys}){}\<[E]%
\ColumnHook
\end{hscode}\resethooks
We will use this relation in the specification of our efficient computation to ensure that the small set of polynomials computed, still ``dominates'' the big set of all the polynomials generated by \ensuremath{\ensuremath{\Varid{genAlg}_{\Varid{n}}}\;\Varid{f}}.

But first we introduce the helper function \ensuremath{\Varid{thin}\mathop{:}\Conid{Set}\;\Varid{a}\,\to\,\Conid{Set}\;\Varid{a}} which aims at removing some elements, while still keeping the minima in the set.
Later we will use the function \ensuremath{\ensuremath{\Varid{genAlgT}_{\!\Varid{n}}}\;\Varid{f}} specified similarly to \ensuremath{\ensuremath{\Varid{genAlg}_{\Varid{n}}}\;\Varid{f}} but using the helper function \ensuremath{\Varid{thin}}.
It has to refine the relation \ensuremath{\Conid{Thin}} which means that if \ensuremath{\Varid{ys}\mathrel{=}\Varid{thin}\;\Varid{xs}} then \ensuremath{\Varid{ys}} must be a subset of \ensuremath{\Varid{xs}} (\ensuremath{\Varid{ys}\subseteq\Varid{xs}}) and \ensuremath{\Varid{ys}} must dominate the rest of \ensuremath{\Varid{xs}} (\ensuremath{\Varid{ys}\mathrel{\ddot{\prec}}(\Varid{xs}\mathbin{\char92 \char92 }\Varid{ys})}).
A trivial (but useless) implementation would be \ensuremath{\Varid{thin}\mathrel{=}\Varid{id}}, and any implementation which removes some ``dominated'' elements could be helpful.
The best we can hope for is that \ensuremath{\Varid{thin}} gives us a set of only incomparable elements.
If \ensuremath{\Varid{thin}} compares all pairs of elements, it can compute a smallest thinning.
In general that may not be needed (and a linear time greedy approximation is good enough), but in some settings almost any algorithmic cost which can reduce the intermediate sets will pay off.
We collect the thinning functions in the type class \ensuremath{\Conid{Thinnable}}:
\begin{hscode}\SaveRestoreHook
\column{B}{@{}>{\hspre}l<{\hspost}@{}}%
\column{3}{@{}>{\hspre}l<{\hspost}@{}}%
\column{5}{@{}>{\hspre}l<{\hspost}@{}}%
\column{18}{@{}>{\hspre}l<{\hspost}@{}}%
\column{28}{@{}>{\hspre}l<{\hspost}@{}}%
\column{E}{@{}>{\hspre}l<{\hspost}@{}}%
\>[B]{}\mathbf{class}\;\Conid{Ord}\;\Varid{a}\Rightarrow \Conid{Thinnable}\;\Varid{a}\;\mathbf{where}{}\<[E]%
\\
\>[B]{}\hsindent{3}{}\<[3]%
\>[3]{}\Varid{thin}{}\<[18]%
\>[18]{}\mathbin{::}\Conid{Set}\;\Varid{a}\,\to\,\Conid{Set}\;\Varid{a}{}\<[E]%
\\
\>[B]{}\hsindent{3}{}\<[3]%
\>[3]{}\Varid{thinStep}{}\<[18]%
\>[18]{}\mathbin{::}\Conid{Set}\;\Varid{a}\,\to\,\Varid{a}\,\to\,\Conid{Set}\;\Varid{a}{}\<[E]%
\\
\>[B]{}\hsindent{3}{}\<[3]%
\>[3]{}\Varid{cmp}{}\<[18]%
\>[18]{}\mathbin{::}{}\<[28]%
\>[28]{}\Varid{a}\,\to\,\Varid{a}\,\to\,\Conid{Maybe}\;\Conid{Ordering}{}\<[E]%
\\
\>[B]{}\hsindent{3}{}\<[3]%
\>[3]{}\Varid{dominatesS}{}\<[18]%
\>[18]{}\mathbin{::}\Conid{Set}\;{}\<[28]%
\>[28]{}\Varid{a}\,\to\,\Varid{a}\,\to\,\mathbb{B}{}\<[E]%
\\[\blanklineskip]%
\>[B]{}\hsindent{3}{}\<[3]%
\>[3]{}\mbox{\onelinecomment  greedy default definitions inspired by \citet{bird_gibbons_2020}}{}\<[E]%
\\
\>[B]{}\hsindent{3}{}\<[3]%
\>[3]{}\Varid{thin}\mathrel{=}\Varid{\Conid{S}.foldl}\;\Varid{thinStep}\;\Conid{S}.\emptyset{}\<[E]%
\\
\>[B]{}\hsindent{3}{}\<[3]%
\>[3]{}\Varid{thinStep}\;\Varid{ys}\;\Varid{x}\mathrel{=}\mathbf{if}\;\Varid{ys}\mathrel{\dot{\prec}}\Varid{x}\;\mathbf{then}\;\Varid{ys}\;\mathbf{else}\;\Varid{\Conid{S}.insert}\;\Varid{x}\;\Varid{ys}{}\<[E]%
\\
\>[B]{}\hsindent{3}{}\<[3]%
\>[3]{}\Varid{ys}\mathrel{\dot{\prec}}\Varid{x}\mathrel{=}\Varid{\Conid{S}.member}\;\T{}\;(\Varid{map}\;(\mathbin{`\Varid{check}`}\Varid{x})\;\Varid{ys}){}\<[E]%
\\
\>[3]{}\hsindent{2}{}\<[5]%
\>[5]{}\mathbf{where}\;\Varid{check}\;\Varid{y}\;\Varid{x}\mathrel{=}\Varid{cmp}\;\Varid{y}\;\Varid{x}\in [\mskip1.5mu \Conid{Just}\;\Conid{LT},\Conid{Just}\;\Conid{EQ}\mskip1.5mu]{}\<[E]%
\ColumnHook
\end{hscode}\resethooks
The greedy \ensuremath{\Varid{thin}} starts from an empty set and considers one element \ensuremath{\Varid{x}} at a time. 
If the set \ensuremath{\Varid{ys}} collected thus far already dominates \ensuremath{\Varid{x}}, it is returned unchanged, otherwise \ensuremath{\Varid{x}} is inserted.
The optimal version also removes from \ensuremath{\Varid{ys}} all elements dominated by~\ensuremath{\Varid{x}}.
It is easy to prove that \ensuremath{\Varid{thin}} implements the specification \ensuremath{\Conid{Thin}}.

The method \ensuremath{\Varid{cmp}} is a more informative version of \ensuremath{(\prec)}: it returns \ensuremath{\Conid{Just}\;\Conid{LT}}, \ensuremath{\Conid{Just}\;\Conid{EQ}}, or \ensuremath{\Conid{Just}\;\Conid{GT}} if the first element is smaller, equal, or greater than the second, respectively, or \ensuremath{\Conid{Nothing}} if they are incomparable. 
%

\paragraph*{Our use of thinning.}
Now we have what we need to specify when an efficient \ensuremath{\ensuremath{\Varid{genAlgT}_{\!\Varid{n}}}\;\Varid{f}} computation is correct.
Our specification (\ensuremath{\Varid{spec}\;\Varid{n}\;\Varid{f}}) states a relation between a (very big) set \ensuremath{\Varid{xs}\mathrel{=}\ensuremath{\Varid{genAlg}_{\Varid{n}}}\;\Varid{f}} and a smaller set \ensuremath{\Varid{ys}\mathrel{=}\ensuremath{\Varid{genAlgT}_{\!\Varid{n}}}\;\Varid{f}} we get by applying thinning at each recursive step.
We want to prove that \ensuremath{\Varid{ys}\subseteq\Varid{xs}} and \ensuremath{\Varid{ys}\mathrel{\ddot{\prec}}(\Varid{xs}\mathbin{\char92 \char92 }\Varid{ys})} because then we know we have kept all the candidates for minimality.
\begin{hscode}\SaveRestoreHook
\column{B}{@{}>{\hspre}l<{\hspost}@{}}%
\column{15}{@{}>{\hspre}l<{\hspost}@{}}%
\column{20}{@{}>{\hspre}l<{\hspost}@{}}%
\column{24}{@{}>{\hspre}l<{\hspost}@{}}%
\column{40}{@{}>{\hspre}l<{\hspost}@{}}%
\column{E}{@{}>{\hspre}l<{\hspost}@{}}%
\>[B]{}\Varid{spec}\;\Varid{n}\;\Varid{f}\mathrel{=}{}\<[15]%
\>[15]{}\mathbf{let}\;{}\<[20]%
\>[20]{}\Varid{xs}{}\<[24]%
\>[24]{}\mathrel{=}\ensuremath{\Varid{genAlg}_{\Varid{n}}}\;{}\<[40]%
\>[40]{}\Varid{f}{}\<[E]%
\\
\>[20]{}\Varid{ys}{}\<[24]%
\>[24]{}\mathrel{=}\ensuremath{\Varid{genAlgT}_{\!\Varid{n}}}\;{}\<[40]%
\>[40]{}\Varid{f}{}\<[E]%
\\
\>[15]{}\mathbf{in}\;{}\<[20]%
\>[20]{}(\Varid{ys}\subseteq\Varid{xs})\mathrel{\wedge}(\Varid{ys}\mathrel{\ddot{\prec}}(\Varid{xs}\mathbin{\char92 \char92 }\Varid{ys})){}\<[E]%
\\[\blanklineskip]%
\>[B]{}\Varid{genAlgT}\mathrel{=}\Varid{genAlgStepThin}\;\Varid{genAlgT}{}\<[E]%
\\
\>[B]{}\Varid{genAlgStepThin}\;\Varid{genT}\;\Varid{n}\;\Varid{f}\mathrel{=}\Varid{thin}\;(\Varid{genAlgStep}\;\Varid{genT}\;\Varid{n}\;\Varid{f}){}\<[E]%
\ColumnHook
\end{hscode}\resethooks
We can first take care of the simplest case (for any \ensuremath{\Varid{n}}).
If the function \ensuremath{\Varid{f}} is constant (returning some \ensuremath{\Varid{b}\mathop{:}\mathbb{B}}), both \ensuremath{\Varid{xs}}
and \ensuremath{\Varid{ys}} will be the singleton set containing \ensuremath{\Varid{res}\;\Varid{b}}.
Thus both properties trivially hold.

We then proceed by induction on \ensuremath{\Varid{n}} to prove \ensuremath{\Conid{S}_{\Varid{n}}\mathrel{=}\forall\,\Varid{f}\mathop{:}\Conid{BoolFun}\;\Varid{n}.\,\,\Varid{spec}\;\Varid{n}\;\Varid{f}}.
In the base case \ensuremath{\Varid{n}\mathrel{=}\mathrm{0}} the function is necessarily constant, and we
have already covered that above.
In the inductive step case, assume the induction hypothesis \ensuremath{\Conid{IH}\mathrel{=}\Conid{S}_{\Varid{n}}} and prove \ensuremath{\Conid{S}_{\Varid{n}\mathbin{+}\mathrm{1}}} for a function \ensuremath{\Varid{f}\mathop{:}\Conid{BoolFun}\;(\Varid{n}\mathbin{+}\mathrm{1})}.
We have already covered the constant function case, so we focus on the main recursive clause of the definitions of \ensuremath{\ensuremath{\Varid{genAlg}_{\Varid{n}}}\;\Varid{f}} and \ensuremath{\ensuremath{\Varid{genAlgT}_{\!\Varid{n}}}\;\Varid{f}} when the fixpoint definitions have been expanded:
\begin{hscode}\SaveRestoreHook
\column{B}{@{}>{\hspre}l<{\hspost}@{}}%
\column{19}{@{}>{\hspre}l<{\hspost}@{}}%
\column{31}{@{}>{\hspre}l<{\hspost}@{}}%
\column{43}{@{}>{\hspre}l<{\hspost}@{}}%
\column{62}{@{}>{\hspre}l<{\hspost}@{}}%
\column{66}{@{}>{\hspre}l<{\hspost}@{}}%
\column{83}{@{}>{\hspre}l<{\hspost}@{}}%
\column{112}{@{}>{\hspre}l<{\hspost}@{}}%
\column{129}{@{}>{\hspre}l<{\hspost}@{}}%
\column{E}{@{}>{\hspre}l<{\hspost}@{}}%
\>[B]{}\ensuremath{\Varid{genAlg}_{\Varid{n}\mathbin{+}\mathrm{1}}}\;{}\<[19]%
\>[19]{}\Varid{f}\mathrel{=}{}\<[31]%
\>[31]{}\{\mskip1.5mu \Varid{pic}\;\Varid{i}\;\Varid{x}_{0}\;{}\<[43]%
\>[43]{}\Varid{x}_{1}\mid \Varid{i}\leftarrow [\mskip1.5mu \mathrm{1}\mathinner{\ldotp\ldotp}\Varid{n}\mskip1.5mu],{}\<[62]%
\>[62]{}\Varid{x}_{0}{}\<[66]%
\>[66]{}\leftarrow \ensuremath{\Varid{genAlg}_{\Varid{n}}}\;{}\<[83]%
\>[83]{}\fixatill{\Varid{i}}{\F{}}{\Varid{f}},\Varid{x}_{1}{}\<[112]%
\>[112]{}\leftarrow \ensuremath{\Varid{genAlg}_{\Varid{n}}}\;{}\<[129]%
\>[129]{}\fixatill{\Varid{i}}{\T{}}{\Varid{f}}\mskip1.5mu\}{}\<[E]%
\\
\>[B]{}\ensuremath{\Varid{genAlgT}_{\!\Varid{n}\mathbin{+}\mathrm{1}}}\;{}\<[19]%
\>[19]{}\Varid{f}\mathrel{=}\Varid{thin}\;{}\<[31]%
\>[31]{}\{\mskip1.5mu \Varid{pic}\;\Varid{i}\;\Varid{y}_{0}\;{}\<[43]%
\>[43]{}\Varid{y}_{1}\mid \Varid{i}\leftarrow [\mskip1.5mu \mathrm{1}\mathinner{\ldotp\ldotp}\Varid{n}\mskip1.5mu],{}\<[62]%
\>[62]{}\Varid{y}_{0}{}\<[66]%
\>[66]{}\leftarrow \ensuremath{\Varid{genAlgT}_{\!\Varid{n}}}\;{}\<[83]%
\>[83]{}\fixatill{\Varid{i}}{\F{}}{\Varid{f}},\Varid{y}_{1}{}\<[112]%
\>[112]{}\leftarrow \ensuremath{\Varid{genAlgT}_{\!\Varid{n}}}\;{}\<[129]%
\>[129]{}\fixatill{\Varid{i}}{\T{}}{\Varid{f}}\mskip1.5mu\}{}\<[E]%
\ColumnHook
\end{hscode}\resethooks
%
%
All subfunctions \ensuremath{\fixatill{\Varid{i}}{\Varid{b}}{\Varid{f}}\mathop{:}\Conid{BoolFun}\;\Varid{n}} used in the recursive calls satisfy the induction hypothesis: \ensuremath{\Varid{spec}\;\Varid{n}\;\fixatill{\Varid{i}}{\Varid{b}}{\Varid{f}}}.
If we name the sets involved in these hypotheses \ensuremath{\fixatill{\Varid{i}}{\Varid{b}}{\Varid{xs}}} and \ensuremath{\fixatill{\Varid{i}}{\Varid{b}}{\Varid{ys}}} we can thus assume \ensuremath{\fixatill{\Varid{i}}{\Varid{b}}{\Varid{ys}}\subseteq\fixatill{\Varid{i}}{\Varid{b}}{\Varid{xs}}} and \ensuremath{\fixatill{\Varid{i}}{\Varid{b}}{\Varid{ys}}\mathrel{\ddot{\prec}}(\fixatill{\Varid{i}}{\Varid{b}}{\Varid{xs}}\mathbin{\char92 \char92 }\fixatill{\Varid{i}}{\Varid{b}}{\Varid{ys}})}.

First, the subset property: we want to prove that \ensuremath{\ensuremath{\Varid{genAlgT}_{\!\Varid{n}\mathbin{+}\mathrm{1}}}\;\Varid{f}\subseteq\ensuremath{\Varid{genAlg}_{\Varid{n}\mathbin{+}\mathrm{1}}}\;\Varid{f}}, or equivalently, \ensuremath{\forall\,\Varid{y}.\,\,(\Varid{y}\in \ensuremath{\Varid{genAlgT}_{\!\Varid{n}\mathbin{+}\mathrm{1}}}\;\Varid{f})\Rightarrow (\Varid{y}\in \ensuremath{\Varid{genAlg}_{\Varid{n}\mathbin{+}\mathrm{1}}}\;\Varid{f})}.
Let \ensuremath{\Varid{y}\in \ensuremath{\Varid{genAlgT}_{\!\Varid{n}\mathbin{+}\mathrm{1}}}\;\Varid{f}}.
We know from the specification of \ensuremath{\Varid{thin}} and the definition of \ensuremath{\ensuremath{\Varid{genAlgT}_{\!\Varid{n}\mathbin{+}\mathrm{1}}}\;\Varid{f}} that \ensuremath{\Varid{y}\mathrel{=}\Varid{pic}\;\Varid{i}\;\Varid{y}_{0}\;\Varid{y}_{1}} for some \ensuremath{\Varid{y}_{0}\in \fixatill{\Varid{i}}{\mathrm{0}}{\Varid{ys}}} and \ensuremath{\Varid{y}_{1}\in \fixatill{\Varid{i}}{\mathrm{1}}{\Varid{ys}}}.
The subset part of the induction hypothesis gives us that \ensuremath{\Varid{y}_{0}\in \fixatill{\Varid{i}}{\mathrm{0}}{\Varid{xs}}} and \ensuremath{\Varid{y}_{1}\in \fixatill{\Varid{i}}{\mathrm{1}}{\Varid{xs}}}.
Thus we can see from the definition of \ensuremath{\ensuremath{\Varid{genAlg}_{\Varid{n}\mathbin{+}\mathrm{1}}}\;\Varid{f}} that \ensuremath{\Varid{y}\in \ensuremath{\Varid{genAlg}_{\Varid{n}\mathbin{+}\mathrm{1}}}\;\Varid{f}}.

Now for the ``domination'' property we need to show that \ensuremath{\forall\,\Varid{x}\in \Varid{xs}\mathbin{\char92 \char92 }\Varid{ys}.\,\,\Varid{ys}\mathrel{\dot{\prec}}\Varid{x}} where \ensuremath{\Varid{xs}\mathrel{=}\ensuremath{\Varid{genAlg}_{\Varid{n}\mathbin{+}\mathrm{1}}}\;\Varid{f}} and \ensuremath{\Varid{ys}\mathrel{=}\ensuremath{\Varid{genAlgT}_{\!\Varid{n}\mathbin{+}\mathrm{1}}}\;\Varid{f}}.
Let \ensuremath{\Varid{x}\in \Varid{xs}\mathbin{\char92 \char92 }\Varid{ys}}.
Given the definition of \ensuremath{\Varid{xs}} it must be of the form \ensuremath{\Varid{x}\mathrel{=}\Varid{pic}\;\Varid{i}\;\Varid{x}_{0}\;\Varid{x}_{1}} where \ensuremath{\Varid{x}_{0}\in \fixatill{\Varid{i}}{\F{}}{\Varid{xs}}} and \ensuremath{\Varid{x}_{1}\in \fixatill{\Varid{i}}{\T{}}{\Varid{xs}}}.
The (second part of the) induction hypothesis provides the existence of \ensuremath{\Varid{y}_{\Varid{b}}\in \fixatill{\Varid{i}}{\Varid{b}}{\Varid{ys}}} such that \ensuremath{\Varid{y}_{\Varid{b}}\prec\Varid{x}_{\Varid{b}}}.
From these \ensuremath{\Varid{y}_{\Varid{b}}} we can build \ensuremath{\Varid{y'}\mathrel{=}\Varid{pic}\;\Varid{i}\;\Varid{y}_{0}\;\Varid{y}_{1}} as a candidate element to ``dominate'' \ensuremath{\Varid{xs}}.

We can now show that \ensuremath{\Varid{y'}\prec\Varid{x}} by polynomial algebra:
\begin{hscode}\SaveRestoreHook
\column{B}{@{}>{\hspre}l<{\hspost}@{}}%
\column{3}{@{}>{\hspre}l<{\hspost}@{}}%
\column{7}{@{}>{\hspre}l<{\hspost}@{}}%
\column{16}{@{}>{\hspre}c<{\hspost}@{}}%
\column{16E}{@{}l@{}}%
\column{17}{@{}>{\hspre}c<{\hspost}@{}}%
\column{17E}{@{}l@{}}%
\column{20}{@{}>{\hspre}l<{\hspost}@{}}%
\column{23}{@{}>{\hspre}l<{\hspost}@{}}%
\column{37}{@{}>{\hspre}l<{\hspost}@{}}%
\column{E}{@{}>{\hspre}l<{\hspost}@{}}%
\>[3]{}\Varid{true}{}\<[E]%
\\
\>[B]{}\implies\mbox{\onelinecomment  Follows from the induction hypothesis}{}\<[E]%
\\
\>[B]{}\hsindent{3}{}\<[3]%
\>[3]{}(\Varid{y}_{0}\prec\Varid{x}_{0}){}\<[17]%
\>[17]{}\mathrel{\wedge}{}\<[17E]%
\>[23]{}(\Varid{y}_{1}\prec\Varid{x}_{1}){}\<[E]%
\\
\>[B]{}\implies\mbox{\onelinecomment  In the interval \ensuremath{(\mathrm{0},\mathrm{1})} both \ensuremath{\mathrm{1}\mathbin{-}\Varid{xP}} and \ensuremath{\Varid{xP}} are positive}{}\<[E]%
\\
\>[B]{}\hsindent{3}{}\<[3]%
\>[3]{}\mathrm{1}\mathbin{+}(\mathrm{1}\mathbin{-}\Varid{xP})\times\Varid{y}_{0}\mathbin{+}\Varid{xP}\times\Varid{y}_{1}~\prec{}\<[37]%
\>[37]{}~\mathrm{1}\mathbin{+}(\mathrm{1}\mathbin{-}\Varid{xP})\times\Varid{x}_{0}\mathbin{+}\Varid{xP}\times\Varid{x}_{1}{}\<[E]%
\\
\>[B]{}\Leftrightarrow\mbox{\onelinecomment  Def. of \ensuremath{\Varid{pic}} for polynomials}{}\<[E]%
\\
\>[B]{}\hsindent{3}{}\<[3]%
\>[3]{}\Varid{pic}\;\Varid{i}\;\Varid{y}_{0}\;\Varid{y}_{1}{}\<[16]%
\>[16]{}\prec{}\<[16E]%
\>[20]{}\Varid{pic}\;\Varid{i}\;\Varid{x}_{0}\;\Varid{x}_{1}{}\<[E]%
\\
\>[B]{}\Leftrightarrow\mbox{\onelinecomment  Def. of \ensuremath{\Varid{y'}} and \ensuremath{\Varid{x}}}{}\<[E]%
\\
\>[B]{}\hsindent{7}{}\<[7]%
\>[7]{}\Varid{y'}\prec\Varid{x}{}\<[E]%
\ColumnHook
\end{hscode}\resethooks
We are not quite done, because \ensuremath{\Varid{y'}} may not be in \ensuremath{\Varid{ys}}.
It is clear from the definition of \ensuremath{\ensuremath{\Varid{genAlgT}_{\!\Varid{n}\mathbin{+}\mathrm{1}}}\;\Varid{f}} that \ensuremath{\Varid{y'}} is in the set \ensuremath{\Varid{ys'}} sent to \ensuremath{\Varid{thin}}, but it may be ``thinned away''.
But, either \ensuremath{\Varid{y'}\in \Varid{ys}\mathrel{=}\Varid{thin}\;\Varid{ys'}} in which case we take the final \ensuremath{\Varid{y}\mathrel{=}\Varid{y'}}, or there exists another \ensuremath{\Varid{y}\in \Varid{ys}} such that \ensuremath{\Varid{y}\prec\Varid{y'}} and then we get get \ensuremath{\Varid{y}\prec\Varid{x}} by transitivity.

To sum up, we have now proved that we can push a powerful \ensuremath{\Varid{thin}} step into the recursive enumeration of all cost polynomials in such a way that any minimum is guaranteed to reside in the much smaller set of polynomials thus computed.

The specific properties we need from \ensuremath{(\prec)} (in addition to the general requirements for thinning) are that \ensuremath{(\Varid{pos}\mathbin{+})} and \ensuremath{(\Varid{pos}\times)} are monotonic (for polynomials \ensuremath{\mathrm{0}\prec\Varid{pos}}) and that \ensuremath{\Varid{q}_{0}\prec\Varid{q}_{1}} implies \ensuremath{\Varid{evalP}\;\Varid{q}_{0}\;\Varid{p}\leq \Varid{evalP}\;\Varid{q}_{1}\;\Varid{p}} for all \ensuremath{\mathrm{0}\leq \Varid{p}\leq \mathrm{1}}.


%
%
%
%
%
%
%



\subsection{Memoization}
\label{sec:memo}
%
The call graph of \ensuremath{\ensuremath{\Varid{genAlgT}_{\!\Varid{n}}}\;\Varid{f}} is the same as the call graph of \ensuremath{\ensuremath{\Varid{genAlg}_{\Varid{n}}}\;\Varid{f}} and, as mentioned above, it can be exponentially big.
Thus, even though thinning helps in making the intermediate sets exponentially smaller, we still have one source of exponential computational complexity to tackle.
Fortunately, the same subfunctions often appear in many different nodes and this means we can save a significant amount of computation time using memoization.

The classical example of memoization is the Fibonacci function.
Naively computing \ensuremath{\Varid{fib}\;(\Varid{n}\mathbin{+}\mathrm{2})\mathrel{=}\Varid{fib}\;(\Varid{n}\mathbin{+}\mathrm{1})\mathbin{+}\Varid{fib}\;\Varid{n}} leads to exponential growth in the number of function calls.
But if we fill in a table indexed by \ensuremath{\Varid{n}} with already computed results we can compute \ensuremath{\Varid{fib}\;\Varid{n}} in linear time.

Similarly, here we ``just'' need to tabulate the result of the calls to \ensuremath{\ensuremath{\Varid{genAlg}_{\Varid{n}}}\;\Varid{f}} so as to avoid recomputation.
The challenge is that the input we need to tabulate is now a Boolean function which is not as nicely structured as a natural number index.
Fortunately, thanks to \citet{DBLP:journals/jfp/Hinze00a}, Elliott, and others we have generic Trie-based memo functions only a hackage library away\footnote{Available on hackage as the \href{https://hackage.haskell.org/package/MemoTrie}{MemoTrie} Haskell package.}.
The \ensuremath{\Conid{MemoTrie}} library provides the \ensuremath{\Conid{Memoizable}} class and suitable instances and helper functions for most types.
We only need to provide a \ensuremath{\Conid{Memoizable}} instance for \ensuremath{\Conid{BDD}}s, and we do this using \ensuremath{\Varid{inSig}} and \ensuremath{\Varid{outSig}} from the \ensuremath{\Conid{BDD}} package (decision-diagrams).
They expose the top-level structure of a \ensuremath{\Conid{BDD}}: \ensuremath{\Conid{Sig}\;\Varid{bf}} is isomorphic to \ensuremath{\Conid{Either}\;\mathbb{B}\;(\Conid{Index},\Varid{bf},\Varid{bf})} where \ensuremath{\Varid{bf}\mathrel{=}\Conid{BDDFun}}.
We define our top-level function \ensuremath{\Varid{genAlgThinMemo}} by applying memoization to \ensuremath{\ensuremath{\Varid{genAlgT}_{\!\Varid{n}}}} (or, more specifcally, to \ensuremath{\Varid{genAlgStepThin}}).

\subsection{Comparing polynomials}
\label{sec:cmp}
As argued in \cref{sec:thin}, the key to an efficient computation of the best cost polynomials is to compare polynomials as soon as possible and throw away those which are ``uniformly worse''.
The specification of \ensuremath{\Varid{p}\prec\Varid{q}} is \ensuremath{\Varid{p}\;\Varid{x}\leq \Varid{q}\;\Varid{x}} for all \ensuremath{\mathrm{0}\leq \Varid{x}\leq \mathrm{1}} and \ensuremath{\Varid{p}\;\Varid{x}\mathbin{<}\Varid{q}\;\Varid{x}} for some \ensuremath{\mathrm{0}\mathbin{<}\Varid{x}\mathbin{<}\mathrm{1}}.
Note that \ensuremath{(\prec)} is a strict pre-order --- if the polynomials cross, neither is ``uniformly worse'' and we keep both.
A simple example of two incomparable polynomials is \ensuremath{\Varid{xP}} and \ensuremath{\mathrm{1}\mathbin{-}\Varid{xP}} which cross at \ensuremath{\Varid{p}\mathrel{=}\nicefrac{1}{2}}.
%

If we have two polynomials \ensuremath{\Varid{p}}, and \ensuremath{\Varid{q}}, we want to know if \ensuremath{\Varid{p}\leq \Varid{q}} for all inputs in the interval \ensuremath{[\mskip1.5mu \mathrm{0},\mathrm{1}\mskip1.5mu]}.
Equivalently, we need to check if \ensuremath{\mathrm{0}\leq \Varid{q}\mathbin{-}\Varid{p}} in that interval.
\begin{hscode}\SaveRestoreHook
\column{B}{@{}>{\hspre}l<{\hspost}@{}}%
\column{E}{@{}>{\hspre}l<{\hspost}@{}}%
\>[B]{}\Varid{cmpPoly}\mathbin{::}(\Conid{Ord}\;\Varid{a},\Conid{Field}\;\Varid{a})\Rightarrow \Conid{Poly}\;\Varid{a}\,\to\,\Conid{Poly}\;\Varid{a}\,\to\,\Conid{Maybe}\;\Conid{Ordering}{}\<[E]%
\\
\>[B]{}\Varid{cmpPoly}\;\Varid{p}\;\Varid{q}\mathrel{=}\Varid{cmpZero}\;(\Varid{q}\mathbin{-}\Varid{p}){}\<[E]%
\ColumnHook
\end{hscode}\resethooks
As the difference is also a polynomial, we can focus our attention to locating polynomial roots in the unit interval.

\begin{figure}[htbp]
    \centering
    \begin{subfigure}[b]{0.3\textwidth}
        \includegraphics[width = \textwidth]{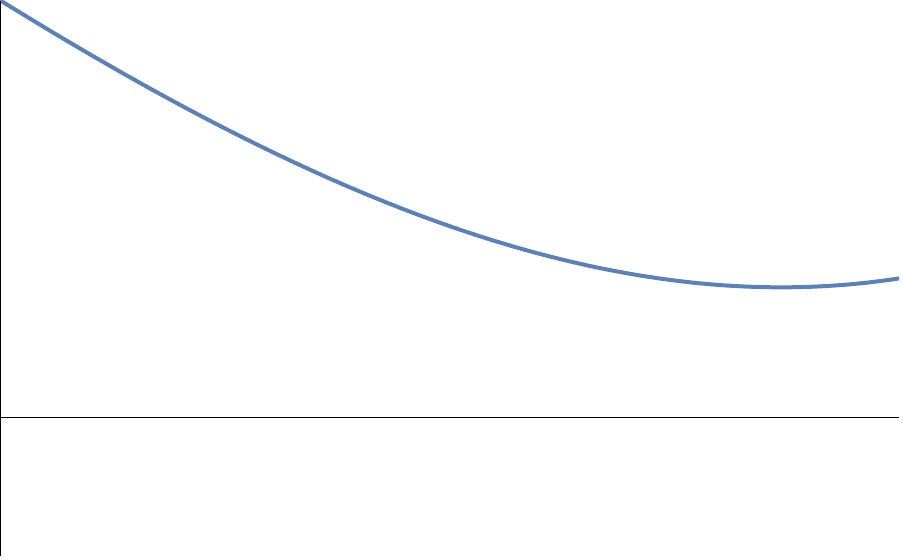}
        \caption{No root.}
        \label{fig:noroot}
    \end{subfigure}
    \hfill
    \begin{subfigure}[b]{0.3\textwidth}
        \includegraphics[width = \textwidth]{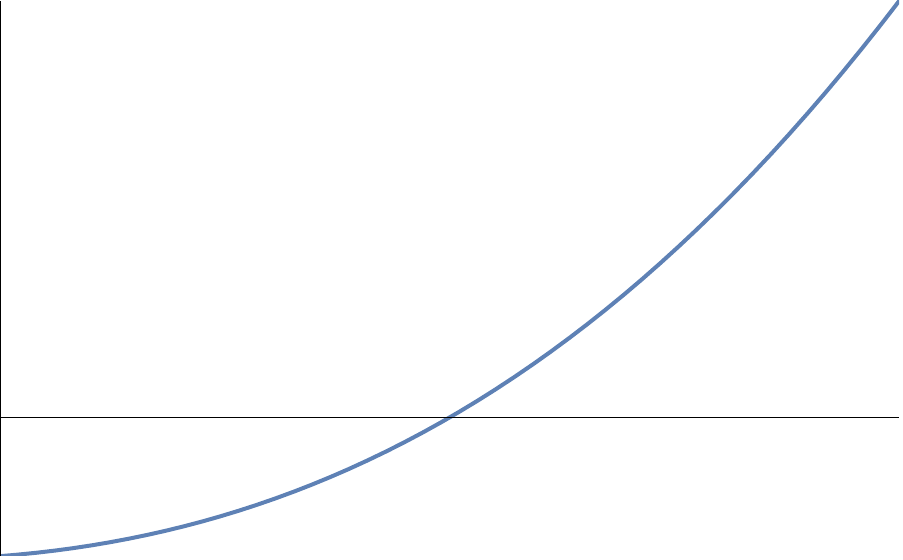}
        \caption{Single root.}
        \label{fig:singleroot}
    \end{subfigure}
    \hfill
    \begin{subfigure}[b]{0.3\textwidth}
        \includegraphics[width = \textwidth]{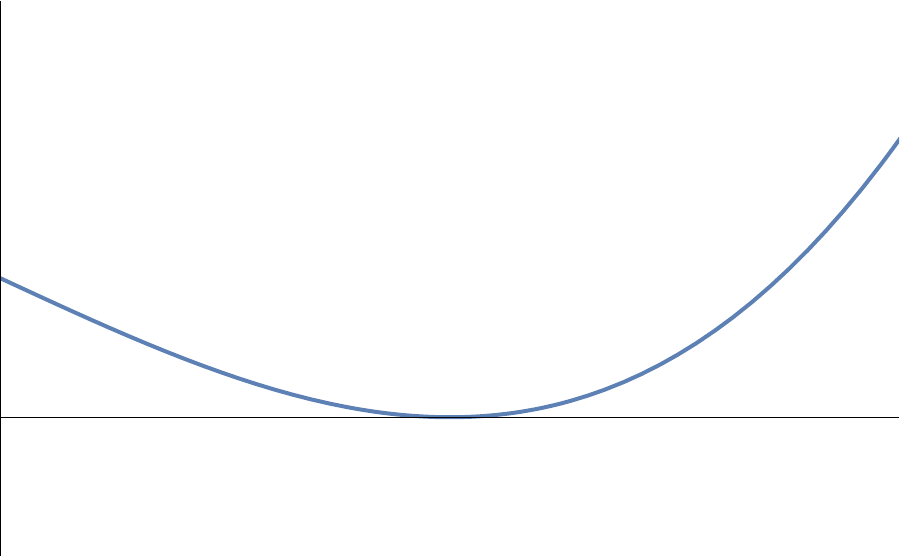}
        \caption{Double root.}
        \label{fig:doubleroot}
    \end{subfigure}
      \caption{To compare two polynomials \ensuremath{\Varid{p}} and \ensuremath{\Varid{q}} we use root counting for \ensuremath{\Varid{q}\mathbin{-}\Varid{p}} and these are the three main cases to consider.}
    \label{fig:roots}
\end{figure}
If there are no roots (\cref{fig:noroot}) in the unit interval, the polynomial stays on ``one side of zero'' and we just need to check the sign of the polynomial at any point.
If there is at least one single-root (\cref{fig:singleroot}), the original polynomials cross and we return \ensuremath{\Conid{Nothing}}.
Similarly for triple-roots or roots of any odd order.
Finally, if the polynomial only has roots of even order (some double-roots, or quadruple-roots, etc.\ as in \cref{fig:doubleroot}) the polynomial stays on one side of zero, and we can check a few points to see what side that is.
(If the number of distinct roots is \ensuremath{\Varid{r}} we check up to \ensuremath{\Varid{r}\mathbin{+}\mathrm{1}} points to make sure at least one will be non-zero and thus tell us on which side of zero the polynomial lies.)

To compare polynomials, we thus need to implement the root-counting functions \ensuremath{\Varid{numRoots}} and \ensuremath{\Varid{numRoots'}}:
\begin{hscode}\SaveRestoreHook
\column{B}{@{}>{\hspre}l<{\hspost}@{}}%
\column{E}{@{}>{\hspre}l<{\hspost}@{}}%
\>[B]{}\Varid{numRoots}\mathbin{::}(\Conid{Ord}\;\Varid{a},\Conid{Field}\;\Varid{a})\Rightarrow \Conid{Poly}\;\Varid{a}\,\to\,\Conid{Int}{}\<[E]%
\\
\>[B]{}\Varid{numRoots}\mathrel{=}\Varid{sum}\mathbin{\circ}\Varid{numRoots'}{}\<[E]%
\\[\blanklineskip]%
\>[B]{}\Varid{numRoots'}\mathbin{::}(\Conid{Ord}\;\Varid{a},\Conid{Field}\;\Varid{a})\Rightarrow \Conid{Poly}\;\Varid{a}\,\to\,[\mskip1.5mu \Conid{Int}\mskip1.5mu]{}\<[E]%
\ColumnHook
\end{hscode}\resethooks
We will not provide all the code here, because that would take us too far from the main topic of the paper, but we will illustrate the main algorithms and concepts for root-counting in \cref{sec:roots}.
The second function computes real root multiplicities: \ensuremath{\Varid{numRoots'}\;\Varid{p}\mathrel{=}[\mskip1.5mu \mathrm{1},\mathrm{3}\mskip1.5mu]} means \ensuremath{\Varid{p}} has one single and one triple root in the open interval \ensuremath{(\mathrm{0},\mathrm{1})}.
From this we get that \ensuremath{\Varid{p}} has \ensuremath{\mathrm{2}\mathrel{=}\Varid{length}\;[\mskip1.5mu \mathrm{1},\mathrm{3}\mskip1.5mu]} distinct real roots and \ensuremath{\mathrm{4}\mathrel{=}\Varid{sum}\;[\mskip1.5mu \mathrm{1},\mathrm{3}\mskip1.5mu]} real roots if we count multiplicities.

Using the root-counting functions, the top-level of the polynomial partial order implementation is as follows:
\begin{hscode}\SaveRestoreHook
\column{B}{@{}>{\hspre}l<{\hspost}@{}}%
\column{3}{@{}>{\hspre}l<{\hspost}@{}}%
\column{10}{@{}>{\hspre}l<{\hspost}@{}}%
\column{12}{@{}>{\hspre}l<{\hspost}@{}}%
\column{18}{@{}>{\hspre}l<{\hspost}@{}}%
\column{42}{@{}>{\hspre}l<{\hspost}@{}}%
\column{57}{@{}>{\hspre}c<{\hspost}@{}}%
\column{57E}{@{}l@{}}%
\column{61}{@{}>{\hspre}l<{\hspost}@{}}%
\column{70}{@{}>{\hspre}l<{\hspost}@{}}%
\column{89}{@{}>{\hspre}l<{\hspost}@{}}%
\column{101}{@{}>{\hspre}l<{\hspost}@{}}%
\column{E}{@{}>{\hspre}l<{\hspost}@{}}%
\>[B]{}\Varid{cmpZero}\mathbin{::}(\Conid{Ord}\;\Varid{a},\Conid{Field}\;\Varid{a})\Rightarrow \Conid{Poly}\;\Varid{a}\,\to\,\Conid{Maybe}\;\Conid{Ordering}{}\<[E]%
\\
\>[B]{}\Varid{cmpZero}\;\Varid{p}{}\<[12]%
\>[12]{}\mid \Varid{isZero}\;\Varid{p}{}\<[57]%
\>[57]{}\mathrel{=}{}\<[57E]%
\>[61]{}\Conid{Just}\;\Conid{EQ}{}\<[E]%
\\
\>[12]{}\mid \Varid{all}\;\Varid{even}\;(\Varid{numRoots'}\;\Varid{p}){}\<[57]%
\>[57]{}\mathrel{=}{}\<[57E]%
\>[61]{}\mathbf{if}\;{}\<[70]%
\>[70]{}\Varid{any}\;(\mathrm{0}\mathbin{<})\;\Varid{vals}\;{}\<[89]%
\>[89]{}\mathbf{then}\;\Conid{Just}\;\Conid{LT}{}\<[E]%
\\
\>[61]{}\mathbf{else}\;\mathbf{if}\;{}\<[70]%
\>[70]{}\Varid{any}\;(\mathrm{0}\mathbin{>})\;\Varid{vals}\;{}\<[89]%
\>[89]{}\mathbf{then}\;\Conid{Just}\;\Conid{GT}{}\<[E]%
\\
\>[61]{}\mathbf{else}\;\Conid{Just}\;\Conid{EQ}{}\<[E]%
\\
\>[12]{}\mid \Varid{otherwise}{}\<[57]%
\>[57]{}\mathrel{=}{}\<[57E]%
\>[61]{}\Conid{Nothing}{}\<[101]%
\>[101]{}\mbox{\onelinecomment  incomparable}{}\<[E]%
\\
\>[B]{}\hsindent{3}{}\<[3]%
\>[3]{}\mathbf{where}\;{}\<[10]%
\>[10]{}\Varid{r}{}\<[18]%
\>[18]{}\mathrel{=}\Varid{length}\;(\Varid{numRoots'}\;\Varid{p}){}\<[42]%
\>[42]{}\mbox{\onelinecomment  the number of distinct roots}{}\<[E]%
\\
\>[10]{}\Varid{rp2}{}\<[18]%
\>[18]{}\mathrel{=}\Varid{fromIntegral}\;(\Varid{r}\mathbin{+}\mathrm{2}){}\<[E]%
\\
\>[10]{}\Varid{points}{}\<[18]%
\>[18]{}\mathrel{=}[\mskip1.5mu \Varid{i}\mathbin{/}\Varid{rp2}\mid \Varid{i}\leftarrow \Varid{take}\;(\Varid{r}\mathbin{+}\mathrm{1})\;(\Varid{iterate}\;(\mathrm{1}\mathbin{+})\;\mathrm{1})\mskip1.5mu]{}\<[E]%
\\
\>[10]{}\Varid{vals}{}\<[18]%
\>[18]{}\mathrel{=}\Varid{map}\;(\Varid{evalP}\;\Varid{p})\;\Varid{points}{}\<[E]%
\ColumnHook
\end{hscode}\resethooks

\subsection{Isolating real roots and Descartes rule of signs}
\label{sec:roots}
This section explains how to do root-counting by combining Yun's algorithm and Descartes rule of signs.
As explained in \cref{sec:cmp} the root-counting is the key to implementing comparison, which is needed for thining.
First out is Yun's algorithm \citep{10.1145/800205.806320} for square-free factorisation: given a polynomial \ensuremath{\Varid{p}} it computes a list of polynomial factors \(p_i\), each of which only has single-roots, and such that \(p = C \prod_{i} {p_i}^i\).
Note the exponent \ensuremath{\Varid{i}}: the factor \ensuremath{\Varid{p}_{2}}, for example, appears squared in \ensuremath{\Varid{p}}.
If \ensuremath{\Varid{p}} only has single-roots, the list from Yun's algorithm has just one element, \ensuremath{\Varid{p}_{1}}, but in any case we get a finite list of polynomials, each of which is ``square-free''.%
\footnote{Yun's algorithm is built around repeated computation of the polynomial greatest common divisor of \ensuremath{\Varid{p}} and its derivative, \ensuremath{\Varid{p'}}.
See the associated code for the details.}

Second in line is Descartes rule of signs which can be used to determine the number of real zeros of a polynomial function.
It tells us that the number of positive real zeros in a polynomial function \ensuremath{\Varid{f}} is the same, or less than by an even number, as the number of changes in the sign of the coefficients.
Together with some polynomial transformations, this is used to count the zeros in the interval \([0,1)\).

If the rule gives zero or one, we are done: we have isolated an interval \([0,1)\) with either no root or exactly one root.
For our use case we don't need to know the actual root, just if it exists in the interval or not.

If the rule gives more than one, we don't quite know the exact number of roots yet (only an upper bound).
In that case we subdivide the interval into the lower \([0,1/2)\) and upper \([1/2, 1)\) halves.
Fortunately the polynomial coefficients can be transformed to make the domain the unit interval again so that we can call ourselves recursively.
After a finite number of steps, this bisection terminates and we get a list of disjoint isolating intervals where we know there is exactly one root in each.
(The number of steps is on the order of the two-logarithm of the
minimum distance between two distinct roots.)

Combining Yun and Descartes, we implement our ``root counter'', and thus our partial order on polynomials.

\section{Results}
Using the method from the previous section we can now calculate the level-$p$-complexity of Boolean functions with our function \ensuremath{\Varid{genAlgThinMemo}}.
First we return to our example from the beginning (\ensuremath{\Varid{sim}_{5}}), where we get several polynomials which are optimal in different intervals.
Then, we calculate the level-$p$-complexity for \ensuremath{\Varid{maj}_{3}^2} which is lower than the proposed result in \citep{jansson2022level}, which means that our current method is better.

\subsection{Level-$p$-complexity for \ensuremath{\Varid{sim}_{5}}}
\label{sec:fAC}
\begin{figure}[htbp]
        \centering
        \includegraphics[width = 0.6\textwidth]{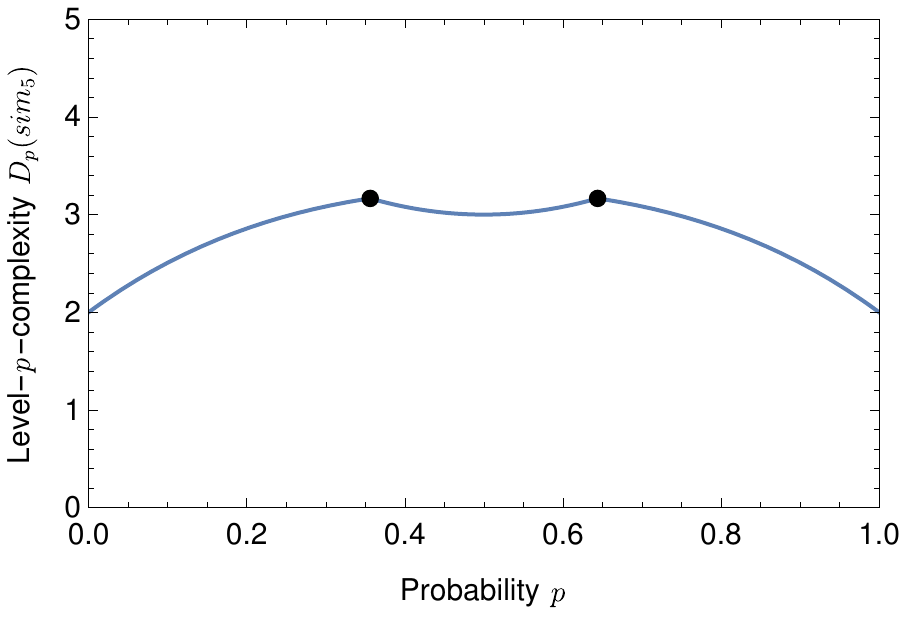}
        \caption{Level-\ensuremath{\Varid{p}}-complexity of \ensuremath{\Varid{sim}_{5}}, where the dots show the intersections of the costs of the decision trees.}
        \label{fig:ACDp}
\end{figure}

When we run \ensuremath{\Varid{genAlgThinMemo}\;\mathrm{5}\;\Varid{sim}_{5}} it returns a set of four polynomials:
\begin{align*}
        \{&P_1(p) = 2 + 6 p - 10 p^2 + 8 p^3 - 4 p^4,  &P_2(p) &= 4 - 2 p - 3 p^2  + 8 p^3 - 2 p^4,\\
        &P_3(p) = 5 - 8 p + 9 p^2 - 2 p^4,&P_4(p) &= 5 - 8 p + 8 p^2 \}
\end{align*}
We don't compute their intersection points, but we know that they do intersect in the unit interval.
The four polynomials were shown already in \cref{fig:4polys}.
The level-$p$-complexity for \ensuremath{\Varid{sim}_{5}} is the piecewise polynomial, pointwise minimum, of these four, with two different polynomials in different intervals: $D_p(\ensuremath{\Varid{sim}_{5}}) = P_4(p)$ for $p \in [\approx0.356,\approx0.644]$ and $D_p(\ensuremath{\Varid{sim}_{5}}) = P_1(p)$ in the rest of the unit interval.
As seen in \cref{fig:ACDp}, the level-$p$-complexity has two maxima.

\subsection{Level-$p$-complexity for \ensuremath{\Varid{maj}_{3}^2}}
\begin{figure}
        \centering
        \includegraphics[width = 0.8\textwidth]{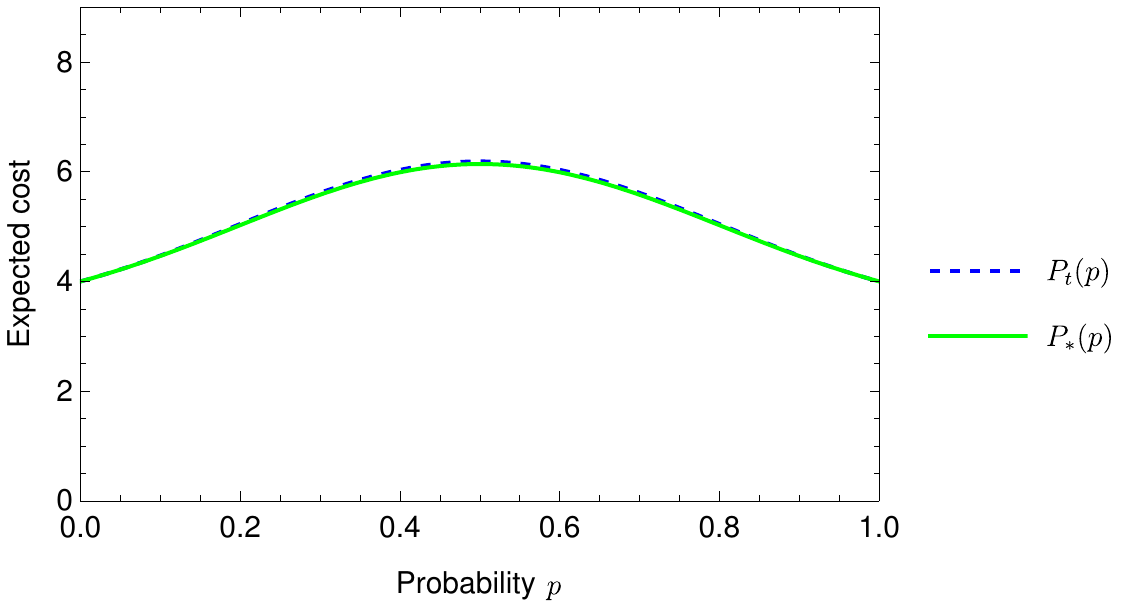}
        \caption{Expected costs of the two different decision trees.
          Because they are very close we also show their difference in \cref{fig:itermajalgsdiff2}.}
        \label{fig:itermajalgs2}
\end{figure}

\begin{figure}
        \centering
        \includegraphics[width = 0.9\textwidth]{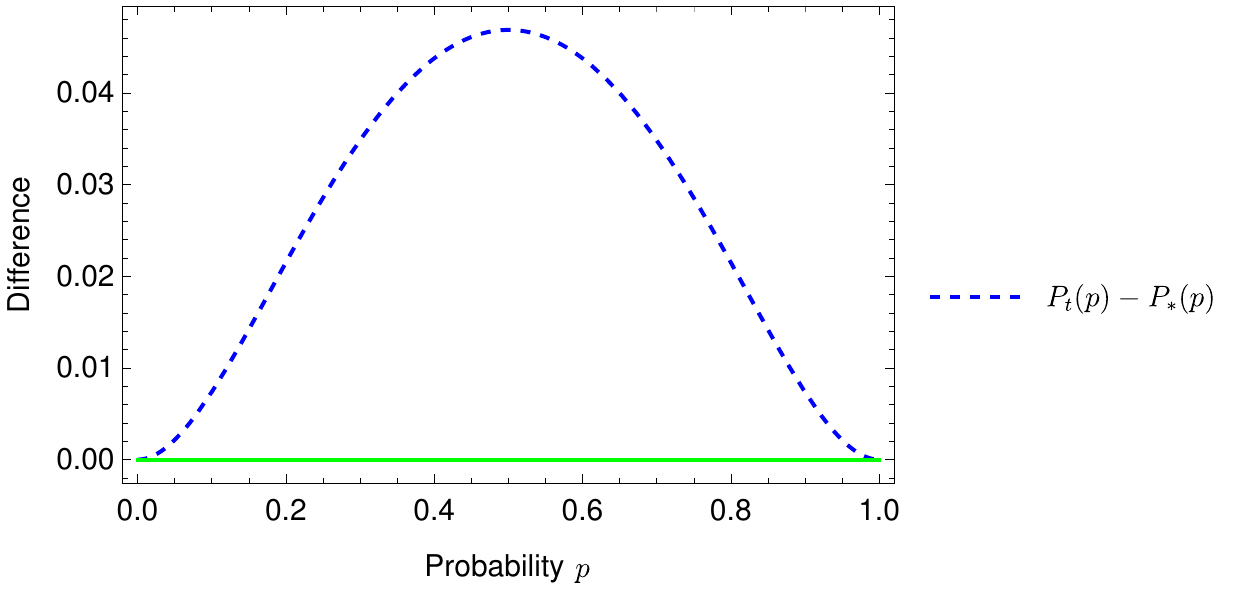}
        \caption{Difference between the expected costs of \ensuremath{\ensuremath{\Conid{T}_{\Varid{t}}}} and \ensuremath{\ensuremath{\Conid{T}_{*}}}.}
        \label{fig:itermajalgsdiff2}
\end{figure}

When running \ensuremath{\Varid{genAlgThinMemo}\;\mathrm{9}\;\Varid{maj}_{3}^2} we get \ensuremath{\{\mskip1.5mu \Conid{P}\;[\mskip1.5mu \mathrm{4},\mathrm{4},\mathrm{6},\mathrm{9},\mathbin{-}\mathrm{61},\mathrm{23},\mathrm{67},\mathbin{-}\mathrm{64},\mathrm{16}\mskip1.5mu]\mskip1.5mu\}}, which means that the expected cost (\(P_*\)) of the best decision tree (\ensuremath{\ensuremath{\Conid{T}_{*}}}) is
$$P_*(p) = 4 + 4 p + 6 p^2 + 9 p^3 - 61 p^4 + 23 p^5 + 67 p^6 - 64 p^7 + 16 p^8\,.$$
This can be compared to the decision tree (that we call \ensuremath{\ensuremath{\Conid{T}_{\Varid{t}}}}) conjectured in \citep{jansson2022level} to be the best.
Its expected cost is slightly higher (thus worse):
$$P_t(p) = 4 + 4 p + 7 p^2 + 6 p^3 - 57 p^4 + 20 p^5 + 68 p^6 - 64 p^7 + 16 p^8\,.$$
The expected costs for decision trees \ensuremath{\ensuremath{\Conid{T}_{*}}} and \ensuremath{\ensuremath{\Conid{T}_{\Varid{t}}}} can be seen in \cref{fig:itermajalgs2}.
Comparing the two polynomials using \ensuremath{\Varid{cmpPoly}\;\ensuremath{\Conid{P}_{*}}\;\ensuremath{\Conid{P}_{\Varid{t}}}} shows that the new one has strictly lower expected cost than the one from the thesis.
The difference, which factors to exactly \(p^2(1-p)^2(1-p+p^2)\), is illustrated in \cref{fig:itermajalgsdiff2}, and we note that it is non-negative in the whole interval.

The value of the polynomials at the endpoints is 4 and the maximum of \ensuremath{\ensuremath{\Conid{P}_{*}}} is $\approx6.14$ compared to the maximum of \ensuremath{\ensuremath{\Conid{P}_{\Varid{t}}}} which is $\approx6.19$.
The conjecture in \citep{jansson2022level} is thus false and the correct formula for the level-$p$-complexity of \ensuremath{\Varid{maj}_{3}^2} is \ensuremath{\ensuremath{\Conid{P}_{*}}}.
At the time of publication of \citep{jansson2022level} it was believed that sifting through all the possible decision trees would be intractable.
Fortunately, using a combination of thinning, memoization, and exact comparison of polynomials, it is now possible to compute the correct complexity in less than a second on the author's laptop.

\section{Conclusions}

This paper describes a Haskell library for computing level-$p$-complexity of Boolean functions, and applies it to two-level iterated majority (\ensuremath{\Varid{maj}_{3}^2}).
The problem specification is straightforward: generate all possible decision trees, compute their expected cost polynomials, and select the best ones.
The implementation is more of a challenge because of two sources of exponential computational cost: an exponential growth in the set of decision trees and an exponential growth in the size of the recursive call graph (the collection of subfunctions).
The library uses thinning to tackle the first and memoization to handle the second source of inefficiency.
In combination with efficient data structures (binary decision diagrams for the Boolean function input, sets of polynomials for the output) this enables computing the level-\ensuremath{\Varid{p}}-complexity for our target example \ensuremath{\Varid{maj}_{3}^2} in less than a second.

From the mathematics point of view the strength of the methods used in this paper to compute the level-\ensuremath{\Varid{p}}-complexity is that we can get a correct result to something which is very hard to calculate by hand.
%
From a computer science point of view the paper is an instructive example of how a combination of algorithmic and symbolic tools can tame a doubly exponential computational cost.

The library uses type-classes for separation of concerns: the actual implementation type for Boolean functions (the input) is abstracted over by the \ensuremath{\Conid{BoFun}} class; and the corresponding type for the output is modelled by the \ensuremath{\Conid{TreeAlg}} class.
We also use our own class \ensuremath{\Conid{Thinnable}} for thinning (and pre-orders), and the \ensuremath{\Conid{Memoizable}} class from hackage.
This means that our main function has the following type:
\begin{hscode}\SaveRestoreHook
\column{B}{@{}>{\hspre}l<{\hspost}@{}}%
\column{20}{@{}>{\hspre}l<{\hspost}@{}}%
\column{E}{@{}>{\hspre}l<{\hspost}@{}}%
\>[B]{}\Varid{genAlgThinMemo}\mathbin{::}{}\<[20]%
\>[20]{}(\Conid{BoFun}\;\Varid{bf},\Conid{Memoizable}\;\Varid{bf},\Conid{TreeAlg}\;\Varid{a},\Conid{Thinnable}\;\Varid{a})\Rightarrow {}\<[E]%
\\
\>[20]{}\mathbb{N}\,\to\,\Varid{bf}\,\to\,\Conid{Set}\;\Varid{a}{}\<[E]%
\ColumnHook
\end{hscode}\resethooks
All the Haskell code is available on GitHub\footnote{The paper repository is at \url{https://github.com/juliajansson/BoFunComplexity}.} and parts of it has been reproduced in Agda to check some of the stronger invariants.
One direction of future work is to complete the Agda formalisation so that we can provide a formally verified library, perhaps helped by \citet{swierstra_2022, 10.1145/3547636}.

The set of polynomials we compute are all incomparable in the pre-order and, together with the thinning relation this means that we actually compute what is called a Pareto front from economics: a set of solutions where no objective can be improved without sacrificing at least one other objective.
It would be interesting to explore this in more detail and to see what the overlap is between thinning as an algorithm design method and different concepts of optimality from economics.

The computed level-$p$-complexity for \ensuremath{\Varid{maj}_{3}^2} is better than the result conjectured in \citep{jansson2022level}, and the library allows easy exploration of other Boolean functions.
With the current library the level-\ensuremath{\Varid{p}}-complexity of iterated majority on 3 levels (27 bits) is out of reach, but with Christian Sattler and Liam Hughes we are exploring a version specialised to ``iterated threshold functions'' which can handle this case (see code in the GitHub repository).

\section*{Acknowledgments}

The authors would like to extend their gratitude to Jeffrey Steif for the idea of exploring level-\ensuremath{\Varid{p}}-complexity and for supervising the preceding work, reported in \citet{jansson2022level}.
Further, we would like to thank Tim Richter and Jeremy Gibbons for taking their time to give valuable feedback on the first draft of this paper.
The authors thank the JFP editors and reviewers, whose helpful and constructive comments have lead to significant improvements of the original manuscript.
The work presented in this paper heavily relies on free software, among others on GHC, Agda, Haskell, git, Emacs, \LaTeX\ and on the Ubuntu operating system, Mathematica, and Visual Studio Code.
It is our pleasure to thank all developers of these excellent products.

\subsection*{Conflicts of Interest}
None.




\label{lastpage01}
\end{document}